\documentclass[10pt,pra,aps,twocolumn,superscriptaddress,floatfix,notitlepage,nofootinbib,reprint]{revtex4-2}
\pdfoutput=1
\usepackage{natbib}
\usepackage{amsmath}
\usepackage{amsfonts}
\usepackage{amssymb}
\usepackage[pdftex]{graphicx}
\usepackage{hyperref}
\usepackage{xcolor}
\usepackage{epsdice}
\usepackage{physics}

\begin{document}
\title{Post-selection and quantum energetics}
\author{Spencer Rogers}
\email{sroge22@ur.rochester.edu}
\affiliation{Department of Physics and Astronomy, University of Rochester, Rochester, NY 14627, USA}
\affiliation{Institute for Quantum Studies, Chapman University, Orange, CA, 92866, USA}

\author{Andrew N. Jordan}
\affiliation{Department of Physics and Astronomy, University of Rochester, Rochester, NY 14627, USA}
\affiliation{Institute for Quantum Studies, Chapman University, Orange, CA, 92866, USA}

\date{\today}
\begin{abstract}
We investigate the anomalous energy change of the measurement apparatus when a qubit is measured in bases that do not commute with energy. We model two possible measurement implementations: one is a quantum clock model with a completely time-independent Hamiltonian, while the other is a Jaynes-Cummings model which is time-dependent but conserves the total excitation number. We look at the mean energy change of the measurement apparatus in both models, conditioned on the qubit post-selection, and find that this change can be much greater than the level spacing of the qubit, like an anomalous weak value. In the clock model, the expression for the apparatus energy shift explicitly contains the weak value of the qubit Hamiltonian. However, in our case, no explicit weak measurements are carried out. Our two models give different results, which we explain to be a consequence of the non-degenerate spectrum of the Jaynes-Cummings model. We compare our calculations in the Jaynes-Cummings model with the experimental data of [J. Stevens, et al, arXiv:2109.09648 (2021)] and find good agreement when the conditions of our derivation are valid.
\end{abstract}
\maketitle

\section{Introduction}
Global energy conservation is the principle that there is an amount of something, called \textit{energy}, that does not change with time. According to this principle, although energy is divided among the subsystems of the world, and distributed differently among these subsystems at different points of time, at all times the total amount of energy is the same. Energy is a shared resource.

There are many complications that arise when attempting to reconcile energy conservation and quantum mechanics. Quantum mechanics introduces \textit{indeterminacy} to physical variables, the notion that these variables may not have a well-defined value prior to being measured. Suppose a small quantum system, a subsystem of the world, has an indeterminate energy. When its energy is measured, and by chance the highest possible value is obtained, does the remainder of the world's energy take its lowest possible value to compensate? The answer turns out to depend on whether the total amount of energy in the world is well-defined or indeterminate. If the total amount of energy is \textit{well-defined}, then when our measured subsystem ends up with its highest possible energy, the remainder of the world ends up with its lowest possible energy, an example of \textit{quantum entanglement}.

The connection between conservation laws, indeterminacy, and entanglement has been noted by various researchers, and a number of results have been found. Most notably, there is the Wigner, Araki, and Yanase (WAY) theorem and its generalizations, which say that perfect measurements (measurements being understood to be related to entanglement in some, albeit mysterious way) of observables that do not commute with additive conserved quantities, like energy, are impossible (see Ref. \cite{wigner1952messung,araki1960measurement,yanase1961optimal,busch2011position,loveridge2011measurement,marvian2012information,ahmadi2013wigner} and Appendix A). While \textit{perfect} measurements of such observables are impossible, very accurate measurements can still be made. These are performed by including an ancillary system with great indeterminacy in the additive conserved quantity  (this indeterminacy serves to reduce the undesirable entanglement ``demanded'' by the conservation law). Similar results hold when the goal is state transformation \cite{popescu2018quantum,popescu2020reference}, as opposed to measurement.  Often, the conserved quantity under consideration is energy. We call the emerging field of research dealing with energy exchange, quantum transformations and measurement, ``quantum energetics.''

\begin{figure}[ht]
\includegraphics[width=\linewidth]{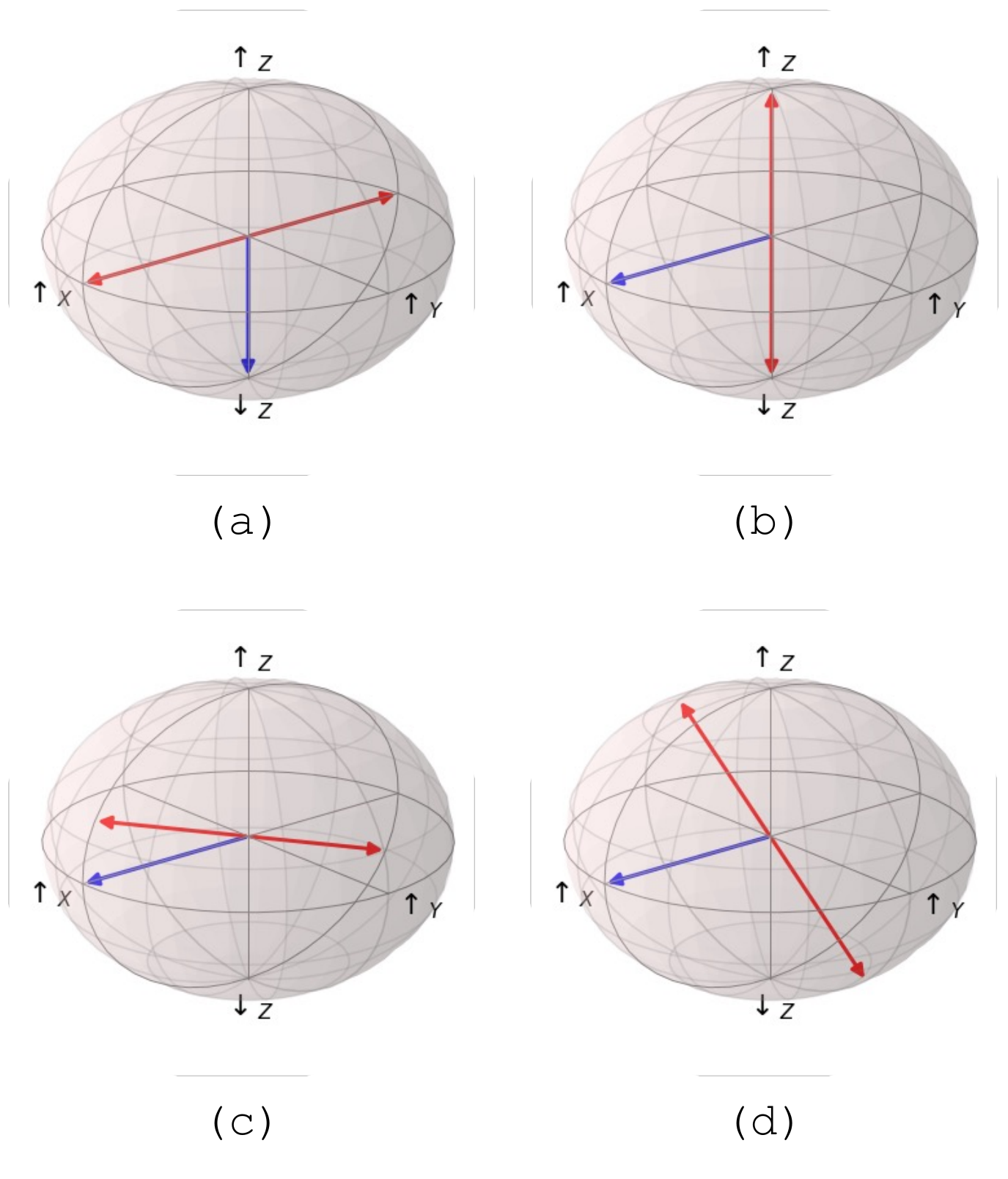}
\caption{(Color online) Examples of different state preparations (single blue arrow) and measurement choices (red axis) are given for a qubit in the Bloch-sphere representation. The excited and ground states are $\ket{\uparrow_z}$ and $\ket{\downarrow_z}$, respectively. (a) The prepared state is an energy eigenstate ($\ket{\downarrow_z}$) and the measurement is of an observable that does not commute with energy ($\hat{\sigma}_x$). (b) The prepared state ($\ket{\uparrow_x}$) has uncertain energy while the measurement is of energy. (c) The qubit initial state and measurement basis are chosen such that the weak value of the qubit Hamiltonian is anomalous (much greater than the level spacing) for one of the measurement outcomes (in particular the unlikely outcome). The example shown features the initial state $\ket{\uparrow_x}$ while the measurement axis lies in the XZ-plane and makes a small angle (exaggerated here for visibility) with respect to the X-axis. (d) The qubit initial state and measurement basis are chosen such that the qubit energy distribution is unchanged, provided that the measurement outcome is not read. The example shown features the prepared state $\ket{\uparrow_x}$, and a measurement axis which lies in the YZ-plane; both measurement outcomes are equally likely. Cases (c) and (d) are both cases where the prepared state is not an eigenstate of energy and the measurement does not commute with energy.} \label{fig:measurementTypes}
\end{figure}

Accurate measurements of observables that do not commute with energy are thus understood to be possible, despite the limitations imposed by conservation of energy. Recent works have focused on utilizing such measurements to power quantum devices, quantum measurement engines \cite{elouard2017extracting,elouard2018efficient,jordan2020quantum,bresque2021two,manikandan2022efficiently,yi2017single,ding2018measurement}. Such works have mostly focused on the energy gained by the measured subsystem itself, as opposed to the energy cost paid by the environment. Of course, if global energy conservation holds, these should balance one another in some way, but the issue becomes quite subtle when we focus, or \textit{post-select}, on particular measurement outcomes. What is the energy change of the remainder of the world, henceforth referred to as the \textit{measurement apparatus} or \textit{environment}, when our measured subsystem is found (post-selected) in a particular state? Previously, Aharanov, Popescu, and Rohrlich considered this question in the particularly striking case of a particle in a box, prepared so as to ``superoscillate'' in a small region faster than any of the Fourier components of its wavefunction \cite{aharonov2021conservation}. When a slit in the box is temporarily opened at the superoscillating region, and the particle happens to escape, the particle's energy is strictly higher than it was previously, raising the question of where the particle got its energy from. The authors concluded that, amazingly, the energy of the box-opener does not change in this case (although the opener's energy does play a role and should have an uncertainty larger than the energy change of the particle). Owing to the paradoxical nature of the effect, Ref. \cite{aharonov2021conservation} went unpublished for over thirty years and sparked numerous supporting and related papers in the interim \cite{berry1994faster,kempf2004unusual,ferreira2006superoscillations,aharonov2011some,rogers2012super,aharonov2017mathematics,berry2017escaping}.

The escaping superoscillation example highlights much about energy conservation with post-selection, but is just one example, and one in which the apparatus seemingly does not supply energy, at that. Indeed, there are many alternative cases in which the measurement apparatus genuinely supplies energy to the measured system (e.g. the quantum measurement engines previously mentioned). The question of how the energy of the measurement apparatus changes when we post-select thus warrants further consideration, and is the focus of our article.

In this article, we take the measured subsystem to be a qubit. This provides a simple regime while still offering insight into surprising effects like the escaping superoscillation. There are various possible choices of the qubit state preparation and measurement basis (see Fig. \ref{fig:measurementTypes}) and one may expect the energetics of the measurement apparatus to vary depending on the choice. We calculate the energy change of the measurement apparatus in each case, to see what effects arise.

This article is structured as follows.

In Section II, we describe two distinct models for strongly measuring a qubit in arbitrary bases. The first is a simple quantum clock model \cite{gisin2018quantum,aharonov2021conservation,aharonov1984quantum,aharonov1998measurement} that utilizes a completely time-independent Hamiltonian, which guarantees energy conservation at the level of unitary evolution. This model was used in the original escaping superoscillation paper, Ref. \cite{aharonov2021conservation}. It was also employed by Gisin and Cruzeiro \cite{gisin2018quantum}, who also studied energy transfer in quantum measurement, but whose paper focused more on the topic of signalling in spin chains \footnote{In particular they examined whether one party, Bob, could gain information about another party Alice's, decision to measure her spin ``by merely looking at the energy change of his [apparatus.]''}. The second model we describe in Section II is a more experimentally relevant Jaynes-Cummings model \cite{haroche2006exploring}. The Hamiltonian here is time-dependent, but commutes with the total excitation number of qubit and oscillator, which stands in for the total energy here. This component of our paper is similar to Ref. \cite{stevens2021energetics}, in which a qubit is driven by radiation, and the photon number change of that radiation, conditioned on the qubit's final energy, is measured.  In both our models, quantum clock and Jaynes-Cummings, the measurement apparatus has its own energy term, whose post-measurement value we can analyze.

In Section III, we give results for the mean energy shift of the measurement apparatus, with post-selection on the state of the qubit. There is a similarity here to the weak value shifts of a noisy pointer in weak measurement, although no weak measurement is explicitly carried out. Interestingly, we find in both models that the mean energy shift of the measurement apparatus may be much larger than the level spacing of the qubit (similar to an anomalous weak value \cite{aharonov1988result,duck1989sense}), an effect essentially witnessed experimentally in Ref. \cite{stevens2021energetics}. We also find that, for the same targeted qubit preparation and post-selection, the two models give \textit{distinct} results for the measurement apparatus energy change; we discuss the reason for this more in Section IV.

In Section IV, we explain why the results of the Jaynes-Cummings model differ from that of the clock model, and discuss the resemblance of the clock model in particular to a deliberate weak measurement of energy. We suggest a possible interpretation of our finding that the weak value of the qubit Hamiltonian occurs in the expression for the clock energy shift. Section V contains our concluding remarks.



\section{Measurement Models}
In this paper, we study two models of qubit measurement. In both models, the measurement apparatus has its own energy term, whose post-measurement value we can evaluate.

\subsection{Quantum clock model}
The quantum clock model \cite{gisin2018quantum,aharonov2021conservation} is the first measurement model we consider. In this three-body model, the total system consists of the measured qubit, a pointer qubit (which stores the measurement outcome), and a quantum clock. The quantum clock is a continuous, one-dimensional system with canonically conjugate variables $\hat{q}$ and $\hat{p}$, and local Hamiltonian $\hat{H}_{clock}=v\hat{p}$. The clock Hamiltonian causes the clock $q$-coordinate to translate with velocity $v$, as can be seen from the commutator $i\comm{\hat{H}_{clock}}{\hat{q}}=v$. The pointer and clock together constitute the \textit{measurement apparatus}. The clock Hamiltonian is the measurement apparatus's energy.

The measured qubit has a Hamiltonian $\hat{H}_0=\frac{\omega_0}{2}\hat{\sigma}_z$ ($\hbar=1$ in this paper), while the pointer qubit has no Hamiltonian of its own, and participates in the dynamics only through the three-body interaction: $\hat{H}_{int}=-\frac{\pi v}{2}\ket{f}\bra{f}\otimes\hat{\sigma}_y^{pointer}\otimes\delta(\hat{q})$, where $\ket{f}$ is a normalized state in the measurement basis of the qubit, the other being $\ket{f_\perp}$. The delta function, $\delta(\hat{q})$, defines a narrow region in $q$-space where the interaction happens, which we call the \textit{interaction region}.

The total Hamiltonian is:

\begin{equation}
\label{eq:clockModelHamiltonian}
\begin{split}
\hat{H}&=\hat{H}_0+\hat{H}_{clock}+\hat{H}_{int}\\*
&=\frac{\omega_0}{2}\hat{\sigma}_z+v\hat{p}-\frac{\pi v}{2}\ket{f}\bra{f}\otimes\hat{\sigma}_y^{pointer}\otimes\delta(\hat{q}).
\end{split}
\end{equation}
The effect of this Hamiltonian is best seen by applying the associated evolution operator to basis kets. Let $\ket{\psi}$ be a state of the qubit. Let $q<0$ so that $\ket{q}$ describes a clock position eigenstate left of the interaction region. For a global state $\ket{\psi}\ket{\downarrow_z^{pointer}}\ket{q}$, evolution then happens in three stages: evolution left of (before) the interaction region (Eq. (\ref{eq:evBefore})), evolution across the interaction region (Eq. (\ref{eq:evAcross})), and evolution right of (after) the interaction region (Eq. (\ref{eq:evAfter})). 

Evolution before the interaction region lasts from time $t=0$ to $t=-q/v$ and is described by:
\begin{equation}\label{eq:evBefore}
\begin{split}
\ket{\psi}\ket{\downarrow_z^{pointer}}\ket{q}\rightarrow e^{+i\hat{H}_0q/v}\ket{\psi}\ket{\downarrow_z^{pointer}}\ket{-\epsilon},\\*
\end{split}
\end{equation}
where $\epsilon$ denotes an infinitesimal distance to the origin. The qubit precesses according to its local Hamiltonian $\hat{H}_0$ for a time $-q/v$, while the clock translates to the left edge of the interaction region.

Evolution across the interaction region occurs at time $t=-q/v$ and is infinitesimally short; it is given by:
\begin{equation}\label{eq:evAcross}
\begin{split}
&e^{+i\hat{H}_0q/v}\ket{\psi}\ket{\downarrow_z^{pointer}}\ket{-\epsilon}\rightarrow\\*
&\Big(\ket{f}\bra{f}e^{+i\hat{H}_0q/v}\ket{\psi}\ket{\uparrow_z^{pointer}}\\*
&+\ket{f_\perp}\bra{f_\perp}e^{+i\hat{H}_0q/v}\ket{\psi}\ket{\downarrow_z^{pointer}}\Big)\ket{+\epsilon}.\\*
\end{split}
\end{equation}
The global state on the left side of the arrow contains a component in which the qubit is in state $\ket{f}$, and a component in which the qubit is in $\ket{f_\perp}$. In the $\ket{f}$ component, the pointer state is flipped from $\ket{\downarrow_z^{pointer}}$ to $\ket{\uparrow_z^{pointer}}$. By contrast, in the $\ket{f_\perp}$ component, the pointer remains in the state $\ket{\downarrow_z}$. Overall, this a controlled-NOT operation, and the result is an entangled superposition of qubit and pointer, characteristic of the standard von Neumann measurement procedure \cite{von2018mathematical}. In fact, Eq. (\ref{eq:evAcross}) is characteristic of the entanglement step of a \textit{perfect} strong measurement of $\ket{f}\bra{f}$, regardless of whether $\ket{f}\bra{f}$ commutes with $\hat{H}_0$. Note, though, that the clock, being in a position eigenstate, has infinite energy uncertainty here, which allows for perfect von Neumann measurement of observables that do not commute with the qubit Hamiltonian (this does not contradict the WAY theorem mentioned in Section I, since energy is infinitely uncertain and thus not meaningfully conserved). Later we will take the clock energy uncertainty to be finite but large, as it would more realistically be.

Evolution right of the interaction region occurs from time $t=-q/v$ to $\infty$. On the right side of the arrow below, we give the state at time $\tau>-q/v$:

\begin{equation}\label{eq:evAfter}
\begin{split}
&\Big(\ket{f}\bra{f}e^{+i\hat{H}_0q/v}\ket{\psi}\ket{\uparrow_z^{pointer}}\\*
&+\ket{f_\perp}\bra{f_\perp}e^{+i\hat{H}_0q/v}\ket{\psi}\ket{\downarrow_z^{pointer}}\Big)\ket{+\epsilon}\rightarrow\\*
&e^{-i\hat{H}_0\left(\tau+q/v\right)}\Big(\ket{f}\bra{f}e^{+i\hat{H}_0q/v}\ket{\psi}\ket{\uparrow_z^{pointer}}\\*
&+\ket{f_\perp}\bra{f_\perp}e^{+i\hat{H}_0q/v}\ket{\psi}\ket{\downarrow_z^{pointer}}\Big)\ket{q+v\tau}\\*
\end{split}
\end{equation}
The time elasped since the clock passed the interaction region is $\tau+q/v$. During this time, the qubit precesses under the action of $\hat{H}_0$, while the clock translates at a rate $v$. There is no further interaction between the bodies.

Eq. (\ref{eq:evBefore})-(\ref{eq:evAfter}) give the evolution of basis kets of the three-body system. In these equations the clock is assumed to be in an eigenstate of $\hat{q}$. In light of the WAY theorem, this is ideal as far as measurement accuracy is concerned (see Appendix A-1 for more on the WAY theorem as applied to the quantum clock model). However, our aim is to assess the energy of the clock (which is the energy of the measurement apparatus, $E_M$), which is impossible in the case of a $\hat{q}$ eigenstate, since energy is infinitely uncertain. To maintain high measurement accuracy while allowing for a meaningful value to be assigned to the shift of the mean energy of the clock, we take the clock to be described at $t=0$ by a narrow \footnote{For a Gaussian of width $\sigma_q$, the energy uncertainty is $\frac{v}{2\sigma_q}$. The \textit{necessary} energy uncertainty for accurate measurement is determined by the qubit preparation and measurement basis, and should be much larger than the quantity in Eq. (\ref{eq:clockMeanEnergyChange}) for \textbf{both} outcomes.} wavefunction $\bra{q}\ket{\phi}$ in $q$-space, localized left of the interaction region. For simplicity, we take this wavefunction to be a single-peaked (at a position $q_0<0$)  and symmetric Gaussian $\bra{q}\ket{\phi}=(2\pi\sigma_q^2)^{-1/4}e^{-(q-q_0)^2/(4\sigma_q^2)}$, but similar results will hold for similar shapes. At $t=0$ the qubit and pointer are in the state $\ket{\psi}\ket{\downarrow_z^{pointer}}$, as before. The complete evolution can then be derived from Eq. (\ref{eq:evBefore})-(\ref{eq:evAfter}) by linearity. The results resemble Eq. (\ref{eq:evBefore})-(\ref{eq:evAfter}) to some degree, owing to the qualitative similarity between a $\hat{q}$-eigenstate and the narrow $\ket{\phi}$. Unlike in Eq. (\ref{eq:evBefore})-(\ref{eq:evAfter}), the clock is no longer perfectly separable from the qubit-pointer subsystem, but weakly entangled, so that its energy depends on the qubit-pointer state. Evolution of the global system from $t=0$ to $\tau\gg-q_0/v$ is given, \textit{approximately}, by:

\begin{equation}\label{eq:evApproximate}
\begin{split}
&\ket{\psi}\ket{\downarrow_z^{pointer}}\ket{\phi}\rightarrow\\*
&\approx e^{-i\hat{H}_0\left(\tau+q_0/v\right)}\ket{f}\bra{f}\ket{i}\ket{\uparrow_z^{pointer}}\ket{\phi_f}\\*
&+e^{-i\hat{H}_0\left(\tau+q_0/v\right)}\ket{f_\perp}\bra{f_\perp}\ket{i}\ket{\downarrow_z^{pointer}}\ket{\phi_\perp}.
\end{split}
\end{equation}
Here we have defined the qubit \textit{pre-selection}, $\ket{i}\equiv e^{+i\hat{H}_0q_0/v}\ket{\psi}$, which is the expected state of the qubit just before the measurement interaction (on account of the qubit and clock states at $t=0$). We have also defined the (normalized) conditioned clock states $\{\ket{\phi_f},\ket{\phi_\perp}\}$; more explicit expressions for these, and derivations of the state evolutions, Eq. (\ref{eq:evBefore})-(\ref{eq:evApproximate}), may be found in Appendix B and C.

The way we think of the post-selection is that the pointer is projectively measured in the $\left\{\ket{\uparrow_z^{pointer}},\ket{\downarrow_z^{pointer}}\right\}$ basis after it has had the chance to flip, with the $\ket{\uparrow_z^{pointer}}$ (resp. $\ket{\downarrow_z^{pointer}}$) outcome corresponding to post-selecting on $\ket{f}$ (resp. $\ket{f_\perp}$). This could in principle be done in isolation from the qubit and clock. Importantly, such a measurement would not be able to transfer  energy (``transferring energy'' meaning taking energy from somewhere and putting it somewhere else) to the qubit-clock subsystem, since there would be no physical interaction with the qubit-clock subsystem; a related and important fact is that such measurement satisfies the \textit{Yanase condition} \cite{yanase1961optimal}, in that the pointer observable commutes with the apparatus energy. If the pointer is found in the state $\ket{\uparrow_z^{pointer}}$, the final joint qubit-clock state will be the state $\approx e^{-i\hat{H}_0\left(\tau+q_0/v\right)}\ket{f}\bra{f}\ket{i}\ket{\phi_f}$, up to a normalization factor. The clock has a final mean energy $\bra{\phi_f}\hat{H}_{clock}\ket{\phi_f}$, which can be compared against the initial mean energy $\bra{\phi}\hat{H}_{clock}\ket{\phi}$. Results are given in Section III. Note that, although energy conservation is built into the model via the time-independent Hamiltonian, there is nothing in the model that requires the \textit{conditioned} clock mean energy change to balance the conditioned qubit mean energy change, i.e. we can have $\Delta\langle{E_M}\rangle_f=\bra{\phi_f}\hat{H}_{clock}\ket{\phi_f}-\bra{\phi}\hat{H}_{clock}\ket{\phi}\neq-\left(\bra{f}\hat{H}_0\ket{f}-\bra{i}\hat{H}_0\ket{i}\right)=-\Delta\langle{E_0}\rangle_f$. We will see in Section III that, in some cases, the disparity between the two is quite large; the post-selection of certain qubit states biases the clock energy change.

Note that the value of the interaction Hamiltonian is essentially zero for the vast majority of time, since the clock has negligible support at $q=0$ for the vast majority of time. Thus the total energy can really be thought of as a sum of the qubit energy and the clock energy. Importantly, energy can be exchanged between these subsystems during the brief moment that the clock passes over $q=0$, so the clock may act as a source of energy for the qubit. More concretely, under unitary evolution alone (i.e. if the pointer is \textit{not} projectively measured), the qubit energy changes in general. The reduced qubit state becomes a mixed state: $\hat{\rho}_{qb}\approx |\bra{f}\ket{i}|^2 e^{-i\hat{H}_0(\tau+q_0/v)}\ket{f}\bra{f}e^{+i\hat{H}_0(\tau+q_0/v)}+|\bra{f_\perp}\ket{i}|^2e^{-i\hat{H}_0(\tau+q_0/v)}\ket{f_\perp}\bra{f_\perp}e^{+i\hat{H}_0(\tau+q_0/v)}$, which has the same exact energy as $|\bra{f}\ket{i}|^2 \ket{f}\bra{f}+|\bra{f_\perp}\ket{i}|^2\ket{f_\perp}\bra{f_\perp}$ and a different energy from the initial qubit state in general. Global energy conservation is guaranteed at the level of this unitary evolution, so we must have mean energy balance between the qubit and apparatus, $\Delta\langle{ E_0}\rangle=-\Delta\langle{E_M}\rangle$, or:

\begin{equation}\label{eq:clockMeanEnergyBalance}
\begin{split}
&\underbrace{|\bra{f}\ket{i}|^2\left(\bra{f}\hat{H}_0\ket{f}-\bra{i}\hat{H}_0\ket{i}\right)}_{P(f)\Delta\langle{E_0}\rangle_f}\\*
&+\underbrace{|\bra{f_\perp}\ket{i}|^2\left(\bra{f_\perp}\hat{H}_0\ket{f_\perp}-\bra{i}\hat{H}_0\ket{i}\right)}_{P(f_\perp)\Delta\langle{E_0}\rangle_\perp}\\*
&=-\Bigg[\underbrace{|\bra{f}\ket{i}|^2\left(\bra{\phi_f}\hat{H}_{clock}\ket{\phi_f}-\bra{\phi}\hat{H}_{clock}\ket{\phi}\right)}_{P(f)\Delta\langle{E_M}\rangle_f}\\*
&+\underbrace{|\bra{f_\perp}\ket{i}|^2\left(\bra{\phi_\perp}\hat{H}_{clock}\ket{\phi_\perp}-\bra{\phi}\hat{H}_{clock}\ket{\phi}\right)}_{P(f_\perp)\Delta\langle{E_M}\rangle_\perp}\Bigg].\\*
\end{split}
\end{equation}

We emphasize again that the pointer could in principle be measured in isolation from the qubit and clock, a process which would involve zero transfer of energy to the qubit-clock system from outside sources. The clock suffices as the qubit's sole energy source (Eq. (\ref{eq:clockMeanEnergyBalance})).

\subsection{Jaynes-Cummings model}
The second model we study is a Jaynes-Cummings model. Given access to dispersive measurements of energy and unitary qubit control, qubit properties besides energy can be measured \cite{elouard2017extracting}. We include the Jaynes-Cummings measurement model for two primary reasons. First, it is more experimentally realistic than the quantum clock model, so the results may be tested on modern platforms. Indeed, the authors of Ref. \cite{stevens2021energetics} essentially performed this test already, albeit in a reduced sense. Second, there is the question of the quantum clock result's (Eq. (\ref{eq:clockMeanEnergyChange})) generality, i.e. of whether the conditioned mean energy shift of the measurement apparatus is independent of apparatus details, depending solely on the qubit preparation and measurement choice . Ultimately, we find that the conditioned mean energy shift of the measurement apparatus \textit{differs} in the two models, which shows that details regarding the measurement implementation affect this quantity. We discuss the factors responsible for this difference more in Section IV.

\begin{figure}[ht]
\includegraphics[width=\linewidth]{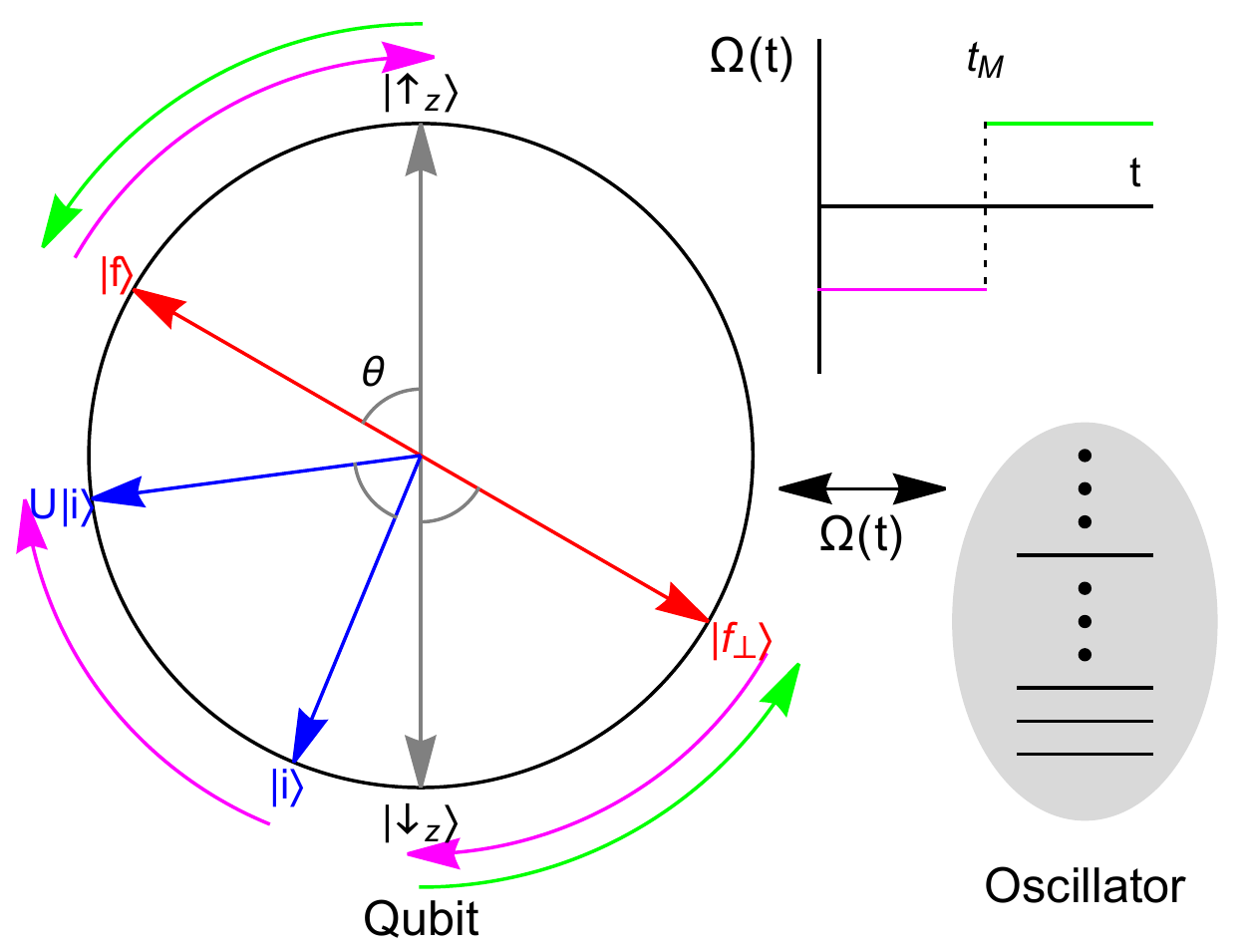}
\caption{(Color online) Energy-conserving measurement protocol using Jaynes-Cummings interaction. We want to measure the observable $\ket{f}\bra{f}$ of the qubit, where $\ket{f}=\cos\frac{\theta}{2}\ket{\uparrow_z}+\sin\frac{\theta}{2}\ket{\downarrow_z}$ ($\ket{\uparrow_z}$ and $\ket{\downarrow_z}$ are the excited and ground states of the qubit), and it is assumed that we have access to projective measurements of the qubit's energy. First (pink, clockwise) the qubit is coupled to an oscillator (in a coherent state with large photon number uncertainty) so as to rotate the qubit by an angle $-\theta$ on the Bloch sphere. This takes $\ket{f}$ to $\ket{\uparrow_z}$ and $\ket{f_\perp}$ to $\ket{\downarrow_z}$. Next, the qubit's energy is measured, at time $t_M$. Lastly (green, counter-clockwise), the qubit-oscillator coupling is reversed, resulting in a qubit rotation of $\theta$. This takes $\ket{\uparrow_z}$ to $\ket{f}$ and $\ket{\downarrow_z}$ to $\ket{f_\perp}$. The qubit-oscillator coupling strength as a function of time, $\Omega(t)$, is plotted and assumed to be piecewise constant over two equal intervals of length $t_M$ to achieve the two rotations. A sample initial qubit state $\ket{i}$ lying in the same Bloch-sphere plane as $\ket{\uparrow_z}$ and $\ket{f}$ is shown in order to demonstrate that the process will in general change the qubit's energy: in the figure, the possible final states (red axis) $\ket{f}$ and $\ket{f_\perp}$ both possess a higher mean value for the qubit Hamiltonian than that of $\ket{i}$, since they are higher on the Bloch-sphere. This does not contradict conservation of energy, since the qubit was allowed to exchange energy with the oscillator. $U\ket{i}$ refers to the approximate qubit state just before the measurement at time $t_M$, assuming the initial qubit state was $\ket{i}$.} \label{fig:JaynesCummingsProtocol}
\end{figure}

In the Jaynes-Cummings model, there is a qubit and a bosonic degree of freedom, which we refer to as the \textit{oscillator}. The two may interact by exchanging excitations. The full Hamiltonian is $\frac{\omega_0}{2}\hat{\sigma}_z+\omega_{osc}\hat{a}^\dagger\hat{a}-i\frac{\Omega(t)}{2}\left(\hat{a}\hat{\sigma}_+-\hat{a}^\dagger\hat{\sigma}_-\right)$, where we allow the last, coupling term to depend explicitly on time, owing to some external controller. If the qubit and oscillator are perfectly on resonance, $\omega_0=\omega_{osc}$ (which is something we assume in this paper), then, in the interaction picture with respect to the bare Hamiltonian $\omega_0\left(\frac{\hat{\sigma}_z}{2}+\hat{a}^\dagger\hat{a}\right)$, the Jaynes-Cummings Hamiltonian is just the interaction term:

\begin{equation}\label{eq:JaynesCummingsHamiltonian}
\hat{H}_{JC}(t)=-i\frac{\Omega(t)}{2}\left(\hat{a}\hat{\sigma}_+-\hat{a}^\dagger\hat{\sigma}_-\right),
\end{equation}
which, importantly, conserves the total excitation number $\hat{n}_T=\ket{\uparrow_z}\bra{\uparrow_z}+\hat{a}^\dagger\hat{a}$. Readers may recognize this interaction term as being responsible for Rabi drive, and wonder why we do not focus on dispersive coupling (the off-resonant case), which is more commonly associated with qubit measurement, instead. The reason is that dispersive coupling allows measurements of the qubit's \textit{energy} ($\hat{\sigma}_z$), whereas we want a model in which observables that do not commute with energy can be measured.

We do not ignore dispersive measurement altogether, though. Rather, we assume that we have access to projective measurements of $\hat{\sigma}_z$, which in practice may be implemented via dispersive coupling to a fresh, off-resonant, probe oscillator without exchanging excitations. Given access to projective measurements of $\hat{\sigma}_z$ and a way of performing rotations about the $Y$ axis of the Bloch sphere, we can effectively measure any $\hat{\Vec{\sigma}}\cdot\Vec{n}$, where $\Vec{n}$ lies in the $XZ$ plane. For example, to measure if the qubit is in the state $\ket{f}$ (assume this is an eigenstate of $\hat{\Vec{\sigma}}\cdot\Vec{n}$), we would perform the $Y$ rotation $\ket{\uparrow_z}\bra{f}+\ket{\downarrow_z}\bra{f_\perp}$ \footnote{That this is a pure $Y$ rotation is validated by the assumption that $\ket{f}$ and $\ket{f_\perp}$ lie in the $XZ$ plane of the Bloch sphere (on opposite poles). Furthermore, we are free to define the phase of $\ket{f_\perp}$ so that this is a pure rotation (as opposed to a pure rotation plus a relative phase shift).}, measure $\hat{\sigma}_z$, and then perform the reverse $Y$ rotation $\ket{f}\bra{\uparrow_z}+\ket{f_\perp}\bra{\downarrow_z}$. This is equivalent to measuring $\hat{\Vec{\sigma}}\cdot\Vec{n}$, since if the measurement result is $\ket{\uparrow_z}$ (resp. $\ket{\downarrow_z}$), the quantum operation is $\ket{f}\bra{f}$ (resp. $\ket{f_\perp}\bra{f_\perp}$). The Jaynes-Cummings interaction, Eq. (\ref{eq:JaynesCummingsHamiltonian}), can approximately generate these $Y$ rotations of the qubit (the rotations are not completely pure; given initial separable states of qubit and oscillator, some entanglement will be generated), and one might notice that it \textit{would be} proportional to $\hat{\sigma}_y$ if $\hat{a}$ and $\hat{a}^\dagger$ were replaced by a real scalar $\alpha$ (a model that employs no bosonic mode, and simply uses $\hat{\sigma}_y$ instead, might describe interaction with a \textit{classical} electromagnetic field). By increasing the uncertainty in $\hat{n}_T$, $Y$ rotations can be performed to arbitrary fidelity (the qubit-oscillator entanglement decreases). This amounts to initializing the oscillator in a coherent state $\ket{\alpha}$ ($\alpha$ real and positive) of sufficiently high photon number $\alpha^2$, and hence photon number uncertainty $\alpha$ (again, the need for such an ancilla is related to the WAY theorem, see Appendix A-2). Then causing the desired $Y$ rotations is a matter of engineering $\Omega(t)$ given $\ket{\alpha}$. For simplicity, we assume $\Omega(t)$ to be piecewise constant in our calculations (see Appendix D for these): a constant $-\Omega_0$ over the first drive (from $0\leq t<t_M$) and the opposite constant $\Omega_0$ over the second (from $t_M<t\leq2t_M$), but other shapes with the same integrations, $\int_0^{t_M}\Omega(t)dt$ and $\int_{t_M}^{2t_M}\Omega(t)dt$, give the same results. The $\hat{\sigma}_z$ measurement occurs between the two drives, instantaneously at time $t_M$. It is worth noting that, if we increase $\alpha$, we also decrease $\Omega_0t_M$ as $1/{\alpha}$ so as to maintain the same rotation angle with more photons.

Let us summarize how this measurement model works (for an illustration, see Fig. \ref{fig:JaynesCummingsProtocol}). The qubit is prepared in a state $\ket{i}$, while the oscillator is prepared in a coherent state $\ket{\alpha}$ of sufficient energy uncertainty. We measure if the qubit is in the state $\ket{f}$, where $\ket{f}$ lies in the $XZ$ plane of the Bloch sphere, by simulating the rotation $\ket{\uparrow_z}\bra{f}+\ket{\downarrow_z}\bra{f_\perp}$ using the Jaynes-Cummings interaction (Eq. (\ref{eq:JaynesCummingsHamiltonian})), then measuring $\hat{\sigma}_z$ directly, and lastly simulating the reverse rotation $\ket{f}\bra{\uparrow_z}+\ket{f_\perp}\bra{\downarrow_z}$ using the Jaynes-Cummings interaction. The Jaynes-Cummings interaction does not change the total excitation number $\hat{n}_T$, and the $\hat{\sigma}_z$ measurement may be performed without exchanging excitations; in this sense, global energy is conserved. Moreover, the qubit and oscillator are allowed to exchange excitations during the Jaynes-Cummings interaction steps, so the oscillator acts as a source of energy for the qubit, which explains how the measurement is capable of changing the qubit's energy in the first place. We can calculate the final mean photon number conditioned on the measurement outcome (post-selection), $\langle{\hat{a}^\dagger\hat{a}}\rangle_{f}$, and compare to the initial mean photon number, $\langle{\hat{a}^\dagger\hat{a}}\rangle_i=\alpha^2$. The results of this analysis are given in the next section.

\begin{figure*}
            \centering            \includegraphics[width=\linewidth]{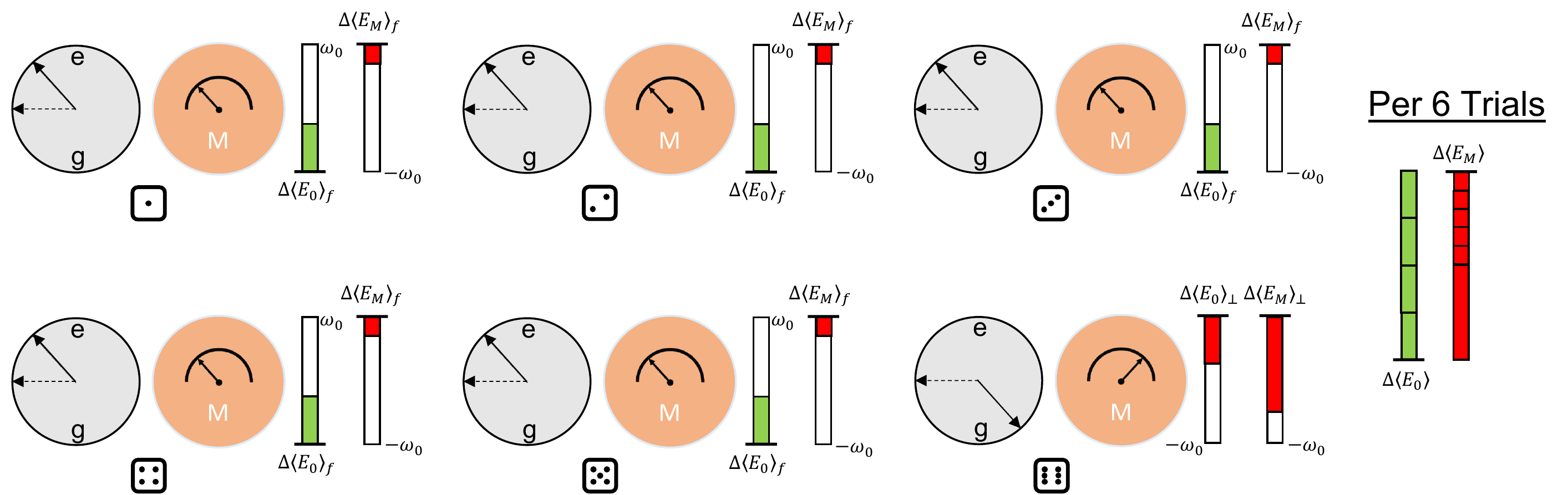}
            \caption{(Color online) Dice-inspired visualization of qubit and measurement apparatus ($M$) energy changes in the clock model: positive changes are shown as shaded green areas (with bar underneath) and negative changes are shown as shaded red areas (with bar above). The qubit pre-selection (dotted arrow on Bloch spheres) is the state of maximal energy uncertainty $\ket{i}=\frac{1}{\sqrt{2}}\left(\ket{e}+\ket{g}\right)$, and we strongly measure the qubit's orientation along an axis that is chosen so as to collapse the qubit onto (solid arrows on Bloch spheres) $\ket{f}=\cos\frac{\theta}{2}\ket{e}+\sin\frac{\theta}{2}\ket{g}$ with probability $5/6$ (\epsdice{1},\epsdice{2},\epsdice{3},\epsdice{4},\epsdice{5}), and $\ket{f_\perp}=\sin\frac{\theta}{2}\ket{e}-\cos\frac{\theta}{2}\ket{g}$ with probability $1/6$ (\epsdice{6}). Here, $\theta=\sin^{-1}\left(\frac{2}{3}\right)$ and $\omega_0$ is the energy difference between the $\ket{e}$ and $\ket{g}$ states. The qubit's mean energy increases with the $\ket{f}$ outcome, $\Delta\langle{E_0}\rangle_f=\omega_0\frac{\sqrt{5}}{6}$, and decreases with the $\ket{f_\perp}$ outcome, $\Delta\langle{E_0}\rangle_\perp=-\omega_0\frac{\sqrt{5}}{6}$. Per \textbf{six} iterations of the measurement, the expected energy gain of the qubit is $\Delta\langle{E_0}\rangle=\omega_0\frac{2\sqrt{5}}{3}$. Global mean energy conservation over all outcomes demands that the measurement apparatus pays for this: $\Delta\langle{E_M}\rangle=-\omega_0\frac{2\sqrt{5}}{3}$. This figure gives an outcome by outcome breakdown of how the apparatus energy changes: $\Delta\langle{E_M}\rangle_f=-\omega_0\frac{1}{3\sqrt{5}}$ and $\Delta\langle{E_M}\rangle_\perp=-\omega_0\frac{\sqrt{5}}{3}$. Notably, in outcome \epsdice{6}, both the qubit and measurement apparatus lose energy. It is similarly possible to choose a measurement basis so that, for some outcome, both qubit and measurement apparatus gain energy.}
            \label{fig:diceExample}
\end{figure*}    

\twocolumngrid

\section{Results}
We now give our results for how the energy of the measurement apparatus changes, depending on the measurement outcome, in each of the two models.

\subsection{Quantum clock model}
In the quantum clock model, the mean energy change of the measurement apparatus, in the pre- and post-selected (PPS) ensemble defined by $\ket{i}$ and $\ket{f}$, takes the simple form:

\begin{equation}\label{eq:clockMeanEnergyChange}
\begin{split}
\Delta\langle{E_M}\rangle_f\approx\bra{\phi_f}\hat{H}_{clock}\ket{\phi_f}-\bra{\phi}\hat{H}_{clock}\ket{\phi}\\*
=\Re\left(\frac{\bra{f}\hat{H}_0\ket{i}}{\bra{f}\ket{i}}\right)-\bra{f}\hat{H}_0\ket{f}.
\end{split}
\end{equation}
The approximation warrants further explanation. $\Delta\langle{E_M}\rangle_f$ asymptotically approaches the value in Eq. (\ref{eq:clockMeanEnergyChange}) as the initial clock energy variance, $\bra{\phi}\hat{H}_{clock}^2\ket{\phi}-\bra{\phi}\hat{H}_{clock}\ket{\phi}^2$, goes to infinity (so that Eq. (\ref{eq:evApproximate}) is valid). This corresponds to making the clock $q$-wavefunction more narrow, and the measurement more accurate. The initial clock energy uncertainty should be large compared to this asymptotic value (over \textbf{both} measurement outcomes) in order for the approximation (and measurement) to be accurate. Interestingly, Eq. (\ref{eq:clockMeanEnergyChange}) contains the weak value \cite{aharonov1988result,duck1989sense} of the qubit Hamiltonian, $\frac{\bra{f}\hat{H}_0\ket{i}}{\bra{f}\ket{i}}$, despite a lack of deliberate weak measurements. Reasons for this are given in Section IV. For a visual example that makes use of Eq. (\ref{eq:clockMeanEnergyChange}), see Fig. \ref{fig:diceExample}. Notably, in some cases the qubit and measurement apparatus both lose (or both gain) energy.

One can check that Eq. (\ref{eq:clockMeanEnergyChange}) is consistent with energy conservation over the complete ensemble of outcomes (Eq. (\ref{eq:clockMeanEnergyBalance})). Eq. (\ref{eq:clockMeanEnergyBalance}) is not particularly restrictive, and in fact allows the conditioned clock mean energy change to exceed the level spacing of the qubit; indeed, by tuning $\ket{i}$ and $\ket{f}$, the weak value term in Eq. (\ref{eq:clockMeanEnergyChange}) may be made arbitrarily large. The energy uncertainty of the clock should also be increased in such extreme cases, to maintain the validity of Eq. (\ref{eq:clockMeanEnergyChange}).

Our result, Eq. (\ref{eq:clockMeanEnergyChange}), has many nice properties, best seen by considering special choices of the qubit preparation and measurement basis. See Fig. \ref{fig:measurementTypes} for Bloch-sphere representations.

\textbf{Qubit prepared in energy eigenstate:} If the qubit is prepared in an energy eigenstate, then the conditioned clock mean energy change is equal and opposite to the conditioned qubit mean energy change: $\Delta\langle{E_M}\rangle_f=\bra{i}\hat{H}_0\ket{i}-\bra{f}\hat{H}_0\ket{f}=-\Delta\langle{E_0}\rangle_f$.

\textbf{Qubit is measured in energy eigenbasis:} If the qubit's energy is measured, then the conditioned clock mean energy change is $\Delta\langle{E_M}\rangle_f=0$.

\textbf{Anomalous weak value:} The conditioned clock mean energy change diverges as $\ket{f}$ and $\ket{i}$ approach orthogonality, provided that $\ket{i}$ has some energy uncertainty. An example is $\ket{i}=\ket{\uparrow_x}$, whereas the post-selection is the unlikely measurement outcome $\ket{f}=\sqrt{(1-\epsilon)/2}\ket{\uparrow_z}-\sqrt{(1+\epsilon)/2}\ket{\downarrow_z}$ ($\epsilon\ll{1}$). The conditioned clock mean energy change is $\Delta\langle{E_M}\rangle_f\approx-\omega_0/\epsilon$, which may be many times larger than the level spacing of the qubit. What saves energy conservation here is that this outcome is rare (probability $P(f)\approx\epsilon^2/4$), and there is a second measurement outcome, $\ket{f_\perp}$, that happens with probability $P(f_\perp)\approx 1-\epsilon^2/4$. With post-selection on $\ket{f_\perp}$, the $\textit{total}$ mean energy change is $\Delta\langle{E_0}\rangle_\perp+\Delta\langle{E_M}\rangle_\perp\approx\epsilon\omega_0/4$.

Global mean energy is conserved over the complete ensemble of outcomes: $P(f)\left(\Delta\langle{E_0}\rangle_f+\Delta\langle{E_M}\rangle_f\right)+P(f_\perp)\left(\Delta\langle{E_0}\rangle_\perp+\Delta\langle{E_M}\rangle_\perp\right)=0$. Since the $\ket{f_\perp}$ contribution scales like $\epsilon$, $P(f)$ scales like $\epsilon^2$, and $\Delta\langle{E_0}\rangle_f$ is bounded by the qubit level spacing, $\Delta\langle{E_M}\rangle_f$ must scale like $1/\epsilon$.

\textbf{Energy non-transferring measurements:} $\ket{i}=\ket{\uparrow_y}$ and the measurement basis lies in the $XZ$ plane of the Bloch sphere. If perfectly executed, such measurements have no effect on the energy probability distribution of the qubit, provided that the measurement outcome is not read (no post-selection). It turns out that, the clock mean energy shift, conditioned on either outcome, is $\Delta\langle{E_M}\rangle_f=0$.

One might expect, given the elegance of Eq. (\ref{eq:clockMeanEnergyChange}) when applied to these cases, that Eq. (\ref{eq:clockMeanEnergyChange}) describes the conditioned mean energy change of any measurement apparatus implementing these qubit measurements. A single counterexample would show that this is not the case, and so now we look at the Jaynes-Cummings model.

\subsection{Jaynes-Cummings model}
In the Jaynes-Cummings model, the photon number change of the oscillator plays the role of the measurement apparatus energy change. It turns out that the resulting expressions for the Jaynes-Cummings model appear somewhat complicated. The clock result (Eq. (\ref{eq:clockMeanEnergyChange})) and Jaynes-Cummings result do, however, share a common form.

\begin{figure}[ht]
    \centering
    \includegraphics[width=\linewidth]{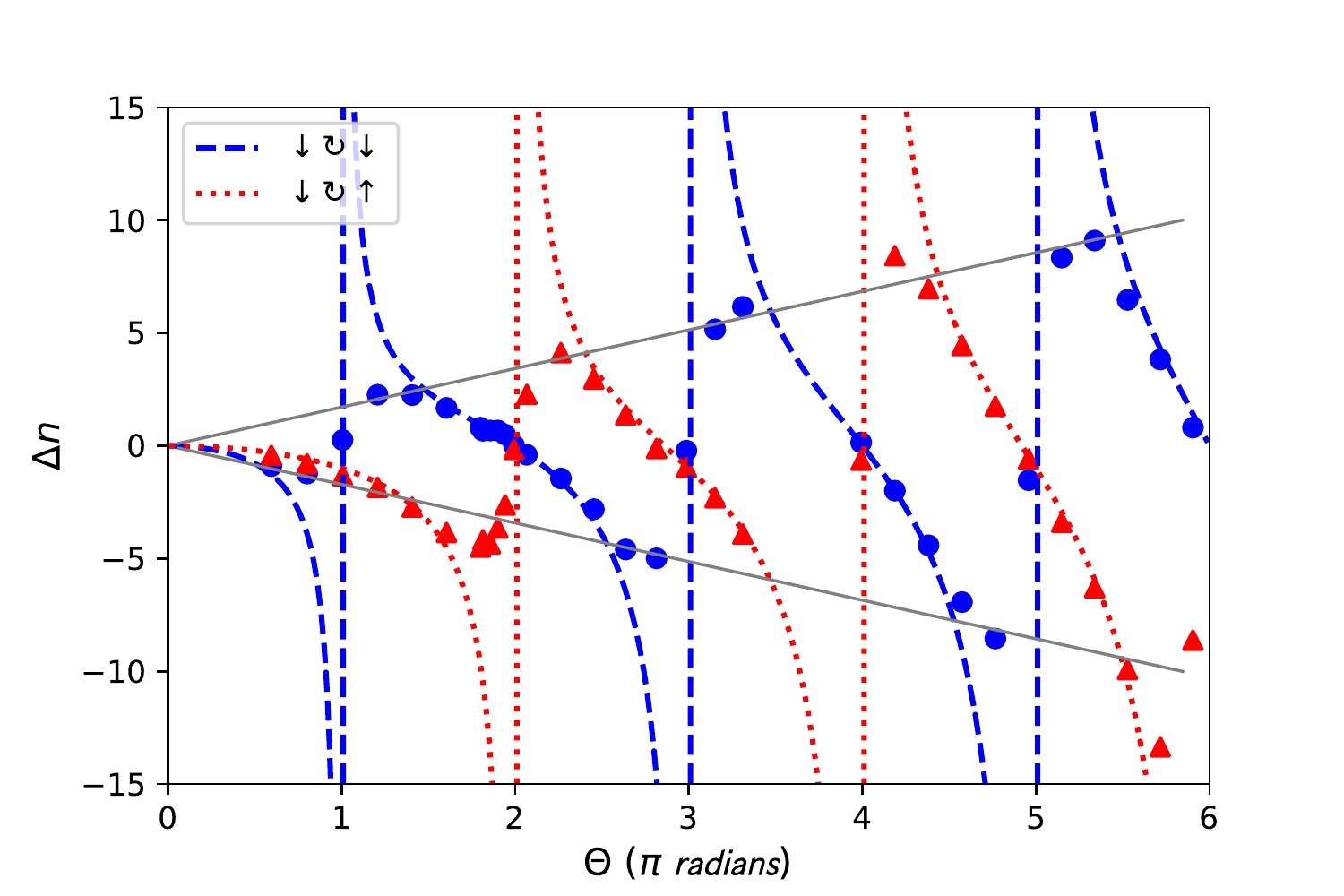}
    \caption{(Color online) Experimental data from Ref. \cite{stevens2021energetics}, Fig. 3. A qubit is resonantly driven from the ground state $\ket{\downarrow_z}$ by an angle $\theta$ around the Bloch sphere, and then measured in the energy basis. The outgoing drive pulse has a photon number shift $\Delta{n}$ relative to the photon number of the initial, incoming pulse. This is measured with post-selection on the ground (blue circles) and excited (red triangles) states.  Dashed-blue and dotted-red curves show theoretical predictions from Eq. (\ref{eq:oscillatorSubChanges}). The data agrees with the theory very well where the theory falls within the linear envelope (solid gray). The envelope scales linearly with the photon number uncertainty $\sqrt{n_{in}}$ of the drive pulse; the way the experiment is performed, $\theta\propto\sqrt{n_{in}}$. Thus the theoretical simplification, Eq. (\ref{eq:oscillatorSubChanges}), holds when the pulse's photon number uncertainty is sufficiently large.}
    \label{fig:data}
\end{figure}

Before giving this form, we introduce the quantities: $\Delta\langle{\tilde{E}_M}\rangle_{\uparrow\circlearrowright\uparrow}$, $\Delta\langle{\tilde{E}_M}\rangle_{\downarrow\circlearrowright\uparrow}$, $\Delta\langle{\tilde{E}_M}\rangle_{\uparrow\circlearrowright\downarrow}$, and $\Delta\langle{\tilde{E}_M}\rangle_{\downarrow\circlearrowright\downarrow}$. For the Jaynes-Cummings model, $\Delta\langle{\tilde{E}_M}\rangle_{\uparrow\circlearrowright\uparrow}$ corresponds to the change in the mean photon number of the oscillator when the qubit is prepared in $\ket{\uparrow_z}$, and the first drive \footnote{This first drive is tuned so as to approximate the unitary rotation of $\ket{f}=\cos\frac{\theta}{2}\ket{\uparrow_z}+\sin\frac{\theta}{2}\ket{\downarrow_z}$ to $\ket{\uparrow_z}$ and hence would approximately take $\ket{\uparrow_z}$ to the ``reflection'' $\ket{f_\leftrightarrow}=\cos\frac{\theta}{2}\ket{\uparrow_z}-\sin\frac{\theta}{2}\ket{\downarrow_z}$.} and $\hat{\sigma}_z$ measurement are performed \textit{and yield} the final state $\ket{\uparrow_z}$. Experiments to measure such quantities have been specifically performed (see Ref. \cite{stevens2021energetics} and Fig. \ref{fig:data}) and agree with the (independently derived) theoretical prediction of the \textbf{Jaynes-Cummings model} that:
\begin{subequations}
\label{eq:oscillatorSubChanges} 
\begin{eqnarray}
\Delta\langle{\tilde{E}_M}\rangle_{\uparrow\circlearrowright\uparrow}(\theta)&=&-\frac{\theta}{2}\tan{\frac{\theta}{2}}, \label{oscillatorSubChangeA}
\\*
\Delta\langle{\tilde{E}_M}\rangle_{\downarrow\circlearrowright\uparrow}(\theta)&=&-1+\frac{\theta}{2}\cot{\frac{\theta}{2}}, \label{oscillatorSubChangeB}
\\*
\Delta\langle{\tilde{E}_M}\rangle_{\uparrow\circlearrowright\downarrow}(\theta)&=&1+\frac{\theta}{2}\cot{\frac{\theta}{2}}, \label{oscillatorSubChangeC}
\\*
\Delta\langle{\tilde{E}_M}\rangle_{\downarrow\circlearrowright\downarrow}(\theta)&=&-\frac{\theta}{2}\tan{\frac{\theta}{2}}. \label{oscillatorSubChangeD}
\end{eqnarray}
\end{subequations}
These results are justified in Appendix D. Here, $-\theta$ is the rotation angle of the first drive (it is worth noting, however, that these functions are even in $\theta$). The purpose of this drive is, ultimately, to help measure the qubit in the basis defined by $\ket{f}=\cos\frac{\theta}{2}\ket{\uparrow_z}+\sin\frac{\theta}{2}\ket{\downarrow_z}$ and $\ket{f_\perp}=-\sin\frac{\theta}{2}\ket{\uparrow_z}+\cos\frac{\theta}{2}\ket{\downarrow_z}$, as one of these two states will be the final state after the second drive (which is by an angle $+\theta$). Whether $\ket{f}$ or $\ket{f_\perp}$ is the final state depends on whether the $\hat{\sigma}_z$ measurement yields $\ket{\uparrow_z}$ or $\ket{\downarrow_z}$, respectively. Keep in mind that there is an underlying approximation: the mean photon number shifts approach the values in Eq. (\ref{eq:oscillatorSubChanges}) as the photon number uncertainty tends to infinity, and larger uncertainties are required to see larger shifts. Interestingly, these photon number shifts are \textbf{not} equal and opposite to the qubit excitation number shifts for the same processes:
\begin{subequations}
\label{eq:qubitEnergyArray} 
\begin{eqnarray}
\Delta\langle{\tilde{E}_0}\rangle_{\uparrow\circlearrowright\uparrow}&=&0, \label{upUpQubitChange}
\\*
\Delta\langle{\tilde{E}_0}\rangle_{\downarrow\circlearrowright\uparrow}&=&+1, \label{downUpQubitChange}
\\*
\Delta\langle{\tilde{E}_0}\rangle_{\uparrow\circlearrowright\downarrow}&=&-1, \label{upDownQubitChange}
\\*
\Delta\langle{\tilde{E}_0}\rangle_{\downarrow\circlearrowright\downarrow}&=&0. \label{downDownQubitChange}
\end{eqnarray}
\end{subequations}
The photon number shifts, Eq. (\ref{eq:oscillatorSubChanges}), may even far exceed the maximum qubit excitation number shift ($1$).

Nevertheless, the qubit and oscillator energy shifts satisfy the conservation relations:

\begin{subequations}
\label{eq:conservationRelations} 
\begin{eqnarray}
P_{\uparrow\circlearrowright\uparrow}\Delta\langle{\tilde{E}_M}\rangle_{\uparrow\circlearrowright\uparrow}+P_{\uparrow\circlearrowright\downarrow}\Delta\langle{\tilde{E}_M}\rangle_{\uparrow\circlearrowright\downarrow}=\label{upConservationRelation}\\*
-\left(P_{\uparrow\circlearrowright\uparrow}\Delta\langle{\tilde{E}_0}\rangle_{\uparrow\circlearrowright\uparrow}+P_{\uparrow\circlearrowright\downarrow}\Delta\langle{\tilde{E}_0}\rangle_{\uparrow\circlearrowright\downarrow}\right),
\nonumber\\*
P_{\downarrow\circlearrowright\uparrow}\Delta\langle{\tilde{E}_M}\rangle_{\downarrow\circlearrowright\uparrow}+P_{\downarrow\circlearrowright\downarrow}\Delta\langle{\tilde{E}_M}\rangle_{\downarrow\circlearrowright\downarrow}=\label{downConservationRelation}\\*
-\left(P_{\downarrow\circlearrowright\uparrow}\Delta\langle{\tilde{E}_0}\rangle_{\downarrow\circlearrowright\uparrow}+P_{\downarrow\circlearrowright\downarrow}\Delta\langle{\tilde{E}_0}\rangle_{\downarrow\circlearrowright\downarrow}\right).\nonumber
\end{eqnarray}
\end{subequations}
These relate the energy changes of the qubit and oscillator when the qubit is prepared in an energy eigenstate ($\ket{\uparrow_z}$ in Eq. (\ref{upConservationRelation}) and $\ket{\downarrow_z}$ in Eq. (\ref{downConservationRelation})), the first drive (by angle $-\theta$ about the $Y$ axis) is applied, and the $\hat{\sigma}_z$ measurement is performed but the outcome is not read. Such processes conserve the total excitation number of qubit and oscillator. The $P_{\uparrow\downarrow\circlearrowright\uparrow\downarrow}$'s are probabilities of the various outcomes, given by:
\begin{subequations}
\label{eq:probabilityArray} 
\begin{eqnarray}
P_{\uparrow\circlearrowright\uparrow}&=&\cos^2\frac{\theta}{2}, \label{upUpProbability}
\\*
P_{\uparrow\circlearrowright\downarrow}&=&\sin^2\frac{\theta}{2}, \label{upDownProbability}
\\*
P_{\downarrow\circlearrowright\uparrow}&=&\sin^2\frac{\theta}{2}, \label{downUpProbability}
\\*
P_{\downarrow\circlearrowright\downarrow}&=&\cos^2\frac{\theta}{2}. \label{downDownProbability}
\end{eqnarray}
\end{subequations}
After the initial drive and $\hat{\sigma}_z$ measurement, the second drive occurs. This drive takes the state $\ket{\uparrow_z}$ to $\ket{f}$ and the state $\ket{\downarrow_z}$ to $\ket{f_\perp}$. Correspondingly, the mean excitation number change of the qubit (which, due to the $\hat{\sigma}_z$ measurement, is in either the state $\ket{\uparrow_z}$ or $\ket{\downarrow_z}$ when the second drive begins) is either $\bra{f}\ket{\uparrow_z}\bra{\uparrow_z}\ket{f}-1$ or $\bra{f_\perp}\ket{\uparrow_z}\bra{\uparrow_z}\ket{f_\perp}$. The mean photon number change of the oscillator from this sub-process is simply equal and opposite to that of the qubit, because during this sub-process there is only the unitary evolution generated by the interaction Hamiltonian $\hat{H}_{JC}$ (Eq. (\ref{eq:JaynesCummingsHamiltonian})), which conserves the total excitation number. Combining this result with Eq. (\ref{eq:oscillatorSubChanges}) gives the total mean photon number change of the oscillator over the entire measurement procedure, provided that the initial qubit state is one of the energy eigenstates, $\ket{\uparrow_z}$ or $\ket{\downarrow_z}$. In general, though, the initial qubit state $\ket{i}$ will be a superposition of the energy eigenstates. Then the total photon number change takes the form:

\begin{equation}
\begin{split}
&\frac{1}{\omega_0}\Delta\langle{E_M}\rangle_f=\frac{1}{|\bra{f}\ket{i}|^2}\Bigg[|\bra{f}\ket{\uparrow_z}\bra{\uparrow_z}\ket{i}|^2\Delta\langle{\tilde{E}_M}\rangle_{\uparrow\circlearrowright\uparrow}\\*
&+|\bra{f}\ket{\downarrow_z}\bra{\downarrow_z}\ket{i}|^2\Delta\langle{\tilde{E}_M}\rangle_{\downarrow\circlearrowright\uparrow}\\*
&+2\Re\left(\bra{f}\ket{\uparrow_z}\bra{\uparrow_z}\ket{i}\overline{\bra{f}\ket{\downarrow_z}}\overline{\bra{\downarrow_z}\ket{i}}\right)\\*
&\left(\frac{\Delta\langle{\tilde{E}_M}\rangle_{\uparrow\circlearrowright\uparrow}+\Delta\langle{\tilde{E}_M}\rangle_{\downarrow\circlearrowright\uparrow}}{2}\right)\Bigg]\\*
&\textcolor{green}{
\underbrace{\textcolor{black}{-\left(\bra{f}\ket{\uparrow_z}\bra{\uparrow_z}\ket{f}-1\right)}}_{{\textcolor{black}{\mathrm{2nd \, drive}}}}
},\\*
\end{split}
\label{eq:MeanEnergyChangeCommonForm}
\end{equation}
for post-selection on $\ket{f}$, and:

\begin{equation}
\begin{split}
&\frac{1}{\omega_0}\Delta\langle{E_M}\rangle_\perp=\frac{1}{|\bra{f_\perp}\ket{i}|^2}\Bigg[|\bra{f_\perp}\ket{\uparrow_z}\bra{\uparrow_z}\ket{i}|^2\Delta\langle{\tilde{E}_M}\rangle_{\uparrow\circlearrowright\downarrow}\\*
&+|\bra{f_\perp}\ket{\downarrow_z}\bra{\downarrow_z}\ket{i}|^2\Delta\langle{\tilde{E}_M}\rangle_{\downarrow\circlearrowright\downarrow}\\*
&+2\Re\left(\bra{f_\perp}\ket{\uparrow_z}\bra{\uparrow_z}\ket{i}\overline{\bra{f_\perp}\ket{\downarrow_z}}\overline{\bra{\downarrow_z}\ket{i}}\right)\\*
&\left(\frac{\Delta\langle{\tilde{E}_M}\rangle_{\uparrow\circlearrowright\downarrow}+\Delta\langle{\tilde{E}_M}\rangle_{\downarrow\circlearrowright\downarrow}}{2}\right)\Bigg]\\*
&\textcolor{green}{
\underbrace{\textcolor{black}{-\bra{f_\perp}\ket{\uparrow_z}\bra{\uparrow_z}\ket{f_\perp}}}_{\textcolor{black}{\mathrm{2nd \, drive}}}
},\\*
\end{split}
\label{eq:MeanEnergyChangeCommonForm2}
\end{equation}
for post-selection on $\ket{f_\perp}$. The first term in Eq. (\ref{eq:MeanEnergyChangeCommonForm}) (resp. Eq. (\ref{eq:MeanEnergyChangeCommonForm2})) takes the form of a weighted average of $\Delta\langle{\tilde{E}_M}\rangle_{\uparrow\circlearrowright\uparrow}$ and $\Delta\langle{\tilde{E}_M}\rangle_{\downarrow\circlearrowright\uparrow}$ (resp. $\Delta\langle{\tilde{E}_M}\rangle_{\uparrow\circlearrowright\downarrow}$ and $\Delta\langle{\tilde{E}_M}\rangle_{\downarrow\circlearrowright\downarrow}$), and contains an interference term which contributes the average of the two. The second terms of Eq. (\ref{eq:MeanEnergyChangeCommonForm}) and (\ref{eq:MeanEnergyChangeCommonForm2}) correspond to the photon number change due to the second drive alone; these terms are emphasized by a \textit{green} underbrace, in keeping with the convention of Fig. \ref{fig:JaynesCummingsProtocol}. Again, there is an underlying approximation to these equations: the mean photon number shifts approach the values claimed here as the photon number uncertainty tends to infinity, and larger uncertainties are required to see larger shifts (for a numerical example, see Fig. A1 of the Appendix).

The \textit{clock} mean energy shift with post-selection, Eq. (\ref{eq:clockMeanEnergyChange}) also follows the form of Eq. (\ref{eq:MeanEnergyChangeCommonForm}) and (\ref{eq:MeanEnergyChangeCommonForm2}). However, different values of the parameters $\Delta\langle{\tilde{E}_M}\rangle_{\uparrow\circlearrowright\uparrow}$, $\Delta\langle{\tilde{E}_M}\rangle_{\downarrow\circlearrowright\uparrow}$, $\Delta\langle{\tilde{E}_M}\rangle_{\uparrow\circlearrowright\downarrow}$, and $\Delta\langle{\tilde{E}_M}\rangle_{\downarrow\circlearrowright\downarrow}$ are needed to fit the equations. For the \textbf{clock model}:
\begin{subequations}
\label{eq:clockSubChanges} 
\begin{eqnarray}
\Delta\langle{\tilde{E}_M}\rangle_{\uparrow\circlearrowright\uparrow}&=&0, \label{clockSubChangeA}
\\*
\Delta\langle{\tilde{E}_M}\rangle_{\downarrow\circlearrowright\uparrow}&=&-1, \label{clockSubChangeB}
\\*
\Delta\langle{\tilde{E}_M}\rangle_{\uparrow\circlearrowright\downarrow}&=&+1, \label{clockSubChangeC}
\\*
\Delta\langle{\tilde{E}_M}\rangle_{\downarrow\circlearrowright\downarrow}&=&0. \label{clockSubChangeD}
\end{eqnarray}
\end{subequations}
There is no drive or intermediate $\hat{\sigma}_z$ measurement in the quantum clock model, so it is not immediately obvious how to interpret the values in Eq. (\ref{eq:clockSubChanges}). Firstly, they are the parameters which fit the clock energy shift to the ``curves'' of Eq. (\ref{eq:MeanEnergyChangeCommonForm}) and (\ref{eq:MeanEnergyChangeCommonForm2}): see Appendix F for proof. Notably, they also satisfy the conservation relations, Eq. (\ref{eq:conservationRelations}). Secondly, it is possible to substitute, in place of $\hat{H}_{JC}$, an alternate qubit-oscillator interaction Hamiltonian, $\hat{H}_D$, which also commutes with the total excitation number but yields the \textit{clock} energy shift instead when applied to the same general protocol (rotate the qubit by $-\theta$, measure $\hat{\sigma}_z$, rotate the qubit by $\theta$). $\Delta\langle{\tilde{E}_M}\rangle_{\uparrow\circlearrowright\uparrow}$ then takes the same meaning as it did for the Jaynes-Cummings model: it corresponds to the change in the mean photon number of the oscillator when the qubit is prepared in $\ket{\uparrow_z}$, and the first drive and $\hat{\sigma}_z$ measurement are performed \textit{and yield} the final state $\ket{\uparrow_z}$. The other values in Eq. (\ref{eq:clockSubChanges}) are analogous. Notice that these values are rather intuitive: they are equal and opposite to the qubit shifts in Eq. (\ref{eq:qubitEnergyArray}). For example, $\Delta\langle{\tilde{E}_M}\rangle_{\downarrow\circlearrowright\uparrow}=-1$ essentially means that when the qubit is prepared in the ground state, driven, and then found in the excited state (thus gaining one quantum of energy overall), the driving source loses one quantum of energy. The values in Eq. (\ref{eq:oscillatorSubChanges}) are rather unintuitive, by contrast. $\hat{H}_D$ is detailed more in Section IV.

\begin{figure}
    \begin{center}
    \includegraphics[width=\linewidth]{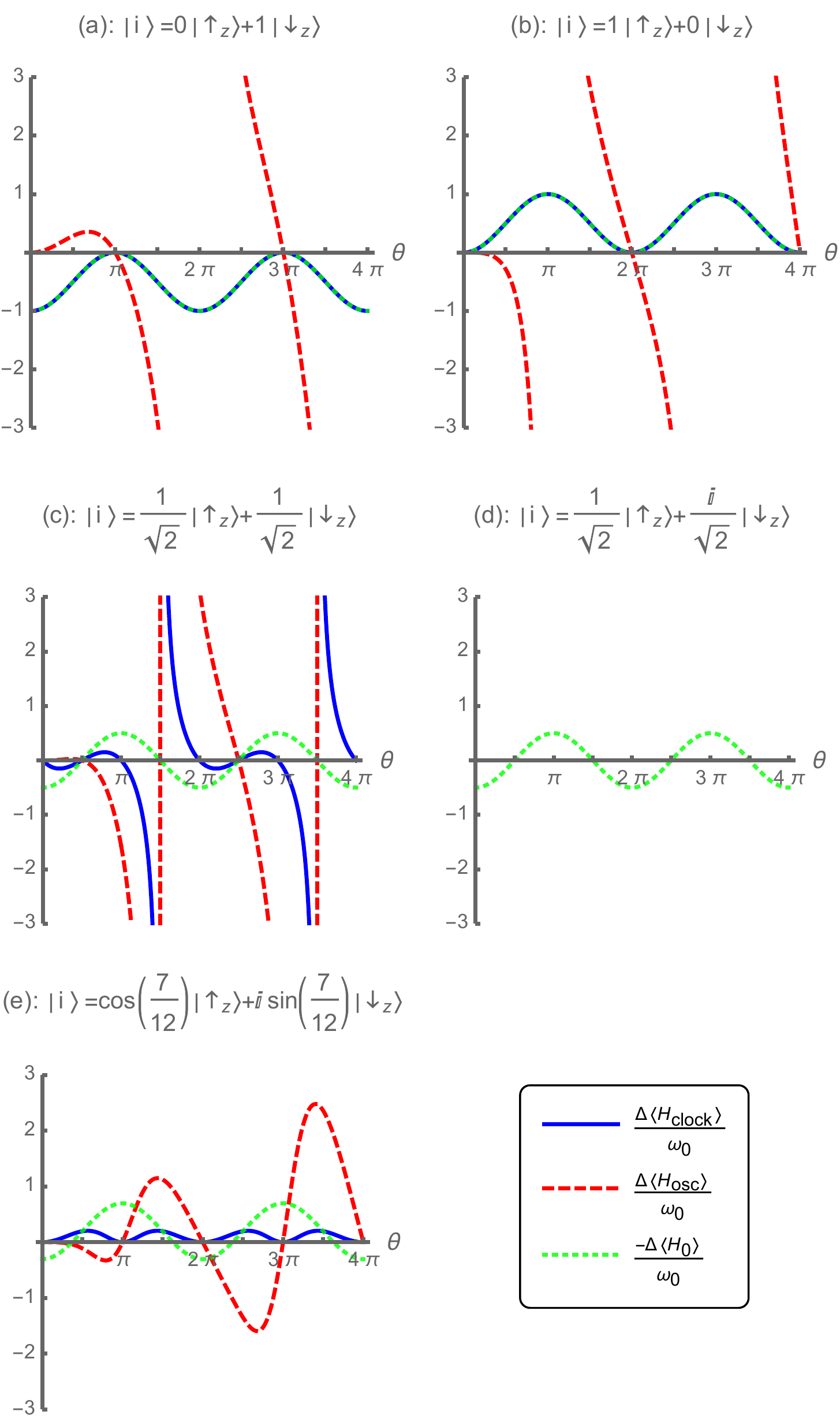}\caption{(Color online) Plots of the conditioned mean energy shift of the measurement apparatus in the clock (solid-blue, Eq. (\ref{eq:clockMeanEnergyChange})) and Jaynes-Cummings (dashed-red, Eq. (\ref{eq:MeanEnergyChangeCommonForm}) with substitutions from Eq. (\ref{eq:oscillatorSubChanges})) models. The negative qubit mean energy change (dotted-green), $\bra{i}\hat{H}_0\ket{i}-\bra{f}\hat{H}_0\ket{f}$, is plotted as well, for reference. In each plot, the initial state $\ket{i}$ of the qubit is fixed, and we sweep through $\theta$, where $\theta$ is the rotation angle used in the Jaynes-Cummings model, and parameterizes the final state (post-selection) according to $\ket{f}=\cos\frac{\theta}{2}\ket{\uparrow_z}+\sin\frac{\theta}{2}\ket{\downarrow_z}$. In (a) and (b), the initial state is an energy eigenstate, and the clock curve matches the negative qubit mean energy change. In (c), the initial state is $\ket{\uparrow_x}$, and both apparatus shifts diverge at $\frac{3\pi}{2}+2n\pi$ (corresponds to $\ket{f}=\ket{\downarrow_x})$. In (d), the initial state is $\ket{\uparrow_y}$, and all possible measurements lie in the $XZ$ plane. Both apparatus shifts are zero. Panel (e) shows a sample state in the $YZ$ plane in order to showcase how the amplitude of the oscillator curve grows with $\theta$, instead of being periodic .}\label{fig:plotGrid}
    \end{center}
\end{figure}

The fact that Eq. (\ref{eq:oscillatorSubChanges}) and (\ref{eq:clockSubChanges}) are different is itself ample evidence that the apparatus energy shifts (with post-selection) are not solely determined by the qubit preparation and targeted measurement basis. The particular measurement implementation is important in determining these conditional shifts. This is further accentuated by the fact that Eq. (\ref{eq:oscillatorSubChanges}) is not purely a function of the targeted measurement basis, $\{\ket{f},\ket{f_\perp}\}$, depending also on the \textbf{complete} drive angle $\theta$ (as opposed to $\theta\mod\pi$) used to measure in that basis (Fig. \ref{fig:plotGrid}e showcases this behavior particularly well).

Given that the clock and Jaynes-Cummings models give different results, we do not expect the oscillator energy changes (with post-selection) to have the same set of nice properties that the clock energy changes had. We consider again special cases of the qubit preparation and measurement basis.

\textbf{Qubit prepared in energy eigenstate}: The conditioned oscillator photon number changes are the values in Eq. (\ref{eq:oscillatorSubChanges}) plus the term corresponding to the second drive (e.g. $\Delta\langle{\tilde{E}_M}\rangle_{\uparrow\circlearrowright\uparrow}(\theta)-\left(\bra{f}\ket{\uparrow_z}\bra{\uparrow_z}\ket{f}-1\right)=-\frac{\theta}{2}\tan\frac{\theta}{2}-\cos\frac{\theta}{2}+1$ for initial state $\ket{\uparrow_z}$ and post-selection $\ket{f}$). $\Delta\langle{E_M}\rangle_f\neq-\Delta\langle{E_0}\rangle_f$, in general, unlike the clock result. See the dashed-red curve, Fig. \ref{fig:plotGrid}a and b.

\textbf{Qubit is measured in energy eigenbasis:} One obvious implementation is to skip the rotations altogether; then there  is only the (unmodelled) $\hat{\sigma}_z$ measurement, and obviously the oscillator's energy does not change. However, the energy of the oscillator can change if one performs unnecessary rotations (as by nonzero multiples of $\pi$), see Fig. \ref{fig:plotGrid}c, in which the dashed-red curve is nonzero at nonzero multiples of $\pi$.

\textbf{Anomalous weak value:} As was the case with the clock, the conditioned oscillator mean energy change tends to diverge as the qubit pre- and post-selection approach orthogonality. Unlike with the clock, this divergence can even be obtained when $\ket{i}$ is an energy eigenstate (see the dashed-red curve in Fig. \ref{fig:plotGrid}a which diverges at $2\pi$ and $4\pi$  and the dashed-red curve in \ref{fig:plotGrid}b which diverges at $\pi$ and $3\pi$).

\textbf{Energy non-transferring measurements:} $\ket{i}=\ket{\uparrow_y}$ and the measurement basis lies in the $XZ$ plane of the Bloch sphere (actually this latter point is assumed throughout this section, without loss of generality since $\ket{i}$ is arbitrary). These measurements have no effect on the energy probability distribution of the qubit, provided that the measurement is accurate and the outcome is not read. Like the clock, the oscillator mean energy change, conditioned on either outcome, is $\Delta\langle{E_M}\rangle_f=0$ (see Fig. \ref{fig:plotGrid}d) \footnote{This statement is true at the level of approximation we have been considering. There may be higher-order effects due to non-ideality of the rotation, but these may be made arbitrarily small by considering higher initial photon numbers.}. The intuition for this is that the $\hat{\sigma}_y$ eigenstate is not affected by the $Y$ drive-rotations.

\section{Discussion}

\subsection{Reconciling the two models}
We now explain why the Jaynes-Cummings model gives different results from the quantum clock model. The answer turns out to have nothing to do with the fact that the Jaynes-Cummings model uses a time-dependent Hamiltonian; rather, the reason has to do with the Jaynes-Cummings interaction's \textit{eigenvalues}.  The eigenvectors of $\hat{H}_{JC}=-i\frac{\Omega}{2}\left(\hat{a}\hat{\sigma}_+-\hat{a}^\dagger\hat{\sigma}_-\right)$ are $\frac{1}{\sqrt{2}}\left(\ket{\uparrow_z,n}\pm i\ket{\downarrow_z,n+1}\right)$, with eigenvalues $\pm\frac{\Omega}{2}\sqrt{n+1}$ \footnote{There is also the ground state eigenvector $\ket{\downarrow_z,0}$, with eigenvalue $0$, but that is unimportant here.}. The resultant qubit oscillation rate thus scales like $\sqrt{n}$; higher photon numbers produce faster rates. We essentially tune $\ket{\alpha}$ and the interaction time $t_M$ such that $\theta\approx\Omega_0{t_M}\sqrt{\bra{\alpha}\hat{a}^\dagger\hat{a}\ket{\alpha}}$, but $\ket{\alpha}$ has a photon number uncertainty $\sqrt{\bra{\alpha}\hat{a}^\dagger\hat{a}\ket{\alpha}}$, and these different photon numbers are out of sync with one another. Put more accurately, $\hat{H}_{JC}$ commutes with the total excitation number $\ket{\uparrow_z}\bra{\uparrow_z}+\hat{a}^\dagger\hat{a}$, and eigenspaces of different total excitation number are out of sync with one another (have different qubit oscillation rates). The probabilities for the $\hat{\sigma}_z$ measurement to yield $\ket{\uparrow_z}$ or $\ket{\downarrow_z}$ depend, weakly, on the eigenspace. This causes the photon number distribution to be modified (in a kind of Bayesian way) depending on the outcome, an example of weak measurement (the present authors do not take credit for this insight, see Ref. \cite{stevens2021energetics} for the same topic). This explains the surprising photon number shifts (Eq. (\ref{eq:oscillatorSubChanges}). This effect on the mean photon number does \textit{not} vanish as we increase the energy uncertainty of $\ket{\alpha}$ (although the qubit rotations may be performed with higher-fidelity this way).

An alternative qubit-oscillator interaction, $\hat{H}_D=-i\frac{\Omega}{2}\left(\hat{L}\hat{\sigma}_+-\hat{L}^\dagger\hat{\sigma}_-\right)$, where $\hat{L}^\dagger=\sum_{n=1}^\infty \ket{n}\bra{n-1}=\sum_{n=0}^\infty \ket{n+1}\bra{n}$, has the same eigenvectors as $\hat{H}_{JC}$, but degenerate eigenvalues $\pm\frac{\Omega}{2}$. In theory, we could implement the same protocol (rotate about $Y$, measure $\hat{\sigma}_z$, then perform the opposite rotation) using a coherent state and this Hamiltonian instead of the Jaynes-Cummings one. Due to the degeneracy of $\hat{H}_D$, the different subspaces of total excitation number would be in sync with one another. The probabilities for the $\hat{\sigma}_z$ measurement to yield $\ket{\uparrow_z}$ or $\ket{\downarrow_z}$ would thus not carry any extra dependence on the subspace. When we perform the same calculation for the conditioned mean energy change of the oscillator using \textit{this} interaction, we get back the clock results (Eq. (\ref{eq:clockMeanEnergyChange}) and (\ref{eq:clockSubChanges})). See Appendix E for details.

It is important to keep in mind that the WAY theorem forbids perfect measurements of observables that do not commute with energy. The clock and Jaynes-Cummings measurement models discussed in this paper only \textit{approximate} perfect measurement in the targeted basis $\{\ket{f},\ket{f_\perp}\}$. These measurement protocols \textit{asymptotically approach} the targeted measurement as the energy uncertainty of the apparatus increases (the initial state $\ket{i}$ also matters), but the approach is different in the two cases, resulting in different limiting behaviors for the conditioned mean energy shift of the measurement apparatus. From our results, it seems apparent that the particular measurement protocol used to approximate the targeted measurement is relevant for the determination of the apparatus energy change (with post-selection). However, one might counter-argue that the clock measurement protocol is, in some way, a more faithful representation of the targeted measurement, and that the Jaynes-Cummings model ultimately describes a slightly different measurement (due to the aforementioned asynchrony between different energy eigenspaces). One must be very clear about what is meant by a ``faithful representation of the targeted measurement," though. If all one requires is fidelity approaching one, then the Jaynes-Cummings measurement model is faithful, since the drive rotations may be performed to arbitrary fidelity by increasing the photon number uncertainty $\sqrt{\langle{n}\rangle}=\alpha$ of the initial coherent state (the product of the vacuum Rabi frequency and interaction time $\Omega_0 t_M$ should simultaneously decrease as $1/\sqrt{\langle{n}\rangle}$ for best results).

\subsection{Presence of the weak value}
We now discuss why we see the weak value appearing in our expression for the conditioned clock mean energy change (Eq. (\ref{eq:clockMeanEnergyChange})), despite a lack of deliberate weak measurements. Note that there are many similarities between the clock measurement model and deliberate weak measurement.

In deliberate weak measurement, the pointer has large quantum uncertainty in some variable (the ``readout'' variable, e.g. $\hat{Q}$). In our case, the apparatus (be it the clock or the oscillator) has a large quantum uncertainty in energy. It must have this for accurate measurement of qubit variables that do not commute with energy, as dictated by the WAY theorem. In deliberate weak measurement, an interaction (e.g. $g(t)\hat{A}\hat{P}$) causes the noisy pointer variable to translate according to the value of the observable $\hat{A}$ of the measured system. The interaction serves as an ``impetus'' for the pointer to shift based on the state of the measured system. If there is no post-selection, the pointer $\hat{Q}$-shift is proportional to the expectation value, $\langle\hat{A}\rangle$. The weak value $\langle\hat{A}\rangle_w$ is observed as the pointer shift conditioned on the result (post-selection) of some later strong measurement. In the clock case, the ``impetus'' is the three-body interaction $\hat{H}_{int}$ (which is tuned for \textit{strong} measurement) combined with the law of conservation of global energy \footnote{As further proof that the law of conservation of global energy also provides this ``impetus,'' consider that if the qubit Hamiltonian were zero, the clock energy shift would vanish. The clock energy shift is linear in the level spacing of the qubit, even though this does not appear in the three-body interaction Hamiltonian, $\hat{H}_{int}$, itself.}. An unread strong measurement of a qubit variable that does not commute with energy generally causes the qubit energy to change, forcing the clock to make up the difference and thus changing the clock's energy. The clock energy shift is, by necessity, equal and opposite to the qubit energy shift (without post-selection); in other words, the clock energy shift \textit{weakly measures} the qubit energy shift caused by an unread strong measurement of some other observable. The weak value-containing expression (Eq. (\ref{eq:clockMeanEnergyChange})) arises when we further condition the clock energy change on the strong measurement outcome.

Note that we could expand our clock model to include a weak energy measurement prior to the strong measurement. This would involve introducing another pointer (this one also with no local Hamiltonian but with a continuous, noisy readout $\hat{Q}$) and another three-body interaction (this one weak, designed to measure $\hat{H}_0$, and between the qubit, clock, and new pointer) positioned so as to happen just before the strong measurement of $\ket{f}\bra{f}$. If we were to condition on the strong measurement outcome $\ket{f}$ and take appropriate limits, the shift of the new pointer would be as though the qubit Hamiltonian took on the (real part of the) weak value, $\Re\left(\frac{\bra{f}\hat{H}_0\ket{i}}{\bra{f}\ket{i}}\right)$, just before the strong measurement interaction.

There is a school of thought that the real part of the weak value, $\Re\left(\frac{\bra{f}\hat{H}_0\ket{i}}{\bra{f}\ket{i}}\right)$, is the \textit{best} description of the qubit energy prior to the strong measurement, given that the qubit is pre-selected in $\ket{i}$ and post-selected in $\ket{f}$ (such talk often pertains to interferometers, where the weak values of path projectors are sometimes attributed a deeper meaning) \cite{hall2004prior,hofmann2021direct,vaidman2013past,danan2013asking,alonso2015can,vaidman2020neutrons}. If we follow this line of thought, then the change in the qubit's energy (due to the strong measurement) is $\bra{f}\hat{H}_0\ket{f}-\Re\left(\frac{\bra{f}\hat{H}_0\ket{i}}{\bra{f}\ket{i}}\right)$, and interestingly, the conditioned clock energy shift (Eq. (\ref{eq:clockMeanEnergyChange})) exactly compensates this. It ``evidences'' that the qubit energy takes on the weak value prior to the strong measurement in essentially the same way that deliberate weak measurements of energy would.

We make no conclusive statement at this time as to whether the Jaynes-Cummings result refutes the reasoning in the previous paragraph. It is apparent that, while a protocol utilizing the Jaynes-Cummings interaction can in some sense measure arbitrary qubit observables to arbitrary accuracy, such a protocol involves the production of seemingly unnecessary (and, arguably, undesirable) correlations between qubit and oscillator (see previous subsection) which may complicate its status as a faithful projective measurement.

\section{Concluding Remarks}
Using two distinct measurement models, the quantum clock model and the Jaynes-Cummings model, we computed the mean energy shift of the measurement apparatus when a measured qubit is found (post-selected) in a particular state. In general, this value is not equal and opposite to the mean energy shift of the qubit. Like the weak value shift of a noisy pointer in weak measurement, the apparatus energy shift may, in both models, be much larger than the level spacing of the qubit. We also identify cases where system and apparatus can both lose (or gain) energy in the post-selected case.  This is consistent with global energy conservation because such events can be traced to energy uncertainty in the initial state of the joint system. While our two models share these similarities, they give different specific results in general, which we explained to be a consequence of the non-degenerate spectrum of the Jaynes-Cummings model, which causes the apparatus's energy change to contain a Bayesian component \cite{stevens2021energetics}. Replacing the Jaynes-Cummings interaction with a similar, degenerate interaction removes this Bayesian component, and returns the clock result. Our work clarifies the principle of energy conservation as applied to measurement in quantum mechanics, and makes a number of interesting experimental predictions, some of which have not been tested yet but are within the grasp of current superconducting qubit technologies (to our knowledge, at the time of writing, the experimentalists behind Ref. \cite{stevens2021energetics} have looked at the quantities in Eq. (\ref{eq:oscillatorSubChanges}), but have not tied them into the broader scheme of Eq. (\ref{eq:MeanEnergyChangeCommonForm}) and (\ref{eq:MeanEnergyChangeCommonForm2})).

\section*{Acknowledgements}
We would like to thank Lea Bresque, Cyril Elouard, and Justin Dressel for helpful discussions. We also thank Jeremy Stevens, Daniel Szombati and Benjamin Huard for agreeing to share their experimental data (Fig. \ref{fig:data} data comes from Ref. \cite{stevens2021energetics}) with us.

This research was funded by the John Templeton Foundation, Grant No. 61835.

\begin{widetext}
\appendix
\renewcommand{\theequation}{A\arabic{equation}}
\renewcommand{\thefigure}{A\arabic{figure}} 

\setcounter{equation}{0}
\setcounter{figure}{0}

\section{WAY theorem}
Our article frequently refers to the Wigner-Arakai-Yanase (WAY) theorem \cite{wigner1952messung,araki1960measurement,yanase1961optimal,busch2011position,loveridge2011measurement,marvian2012information,ahmadi2013wigner}. This result limits the accuracy of measurements that do not commute with additive conserved quantities, such as energy. Perfect measurements of such observables are impossible, but (otherwise) arbitrary accuracy may be achieved by including an ancillary system that exhibits sufficient quantum uncertainty in the additive  conserved quantities. We illustrate here how the WAY theorem relates to the two measurement models in our paper. Readers interested in more rigorous discussions should refer to Ref. \cite{wigner1952messung,araki1960measurement,yanase1961optimal,busch2011position,loveridge2011measurement,marvian2012information,ahmadi2013wigner}.

\subsection{Clock model}
In the quantum clock model, the clock plays the role of the ancillary system with large energy uncertainty, while the qubit observable $\ket{f}\bra{f}$ is measured. Since $\ket{f}$ need not be an energy eigenstate of the qubit, $\ket{f}\bra{f}$ does not commute with energy in general, and WAY theorem limitations apply.

We may consider, at one extreme, that the initial clock state is a position eigenket $\ket{q_0}$, where $q_0<0$. Such a clock state has infinite energy uncertainty, since the clock Hamiltonin is $v\hat{p}$. Concatenating the evolutions in Eq. (\ref{eq:evBefore})-(\ref{eq:evAfter}), and defining the qubit state $\ket{i}=e^{+i\hat{H}_0q_0/v}\ket{\psi}$ at the instant the measurement interaction occurs, gives:

\begin{equation}\label{appendixEq:evIdeal}
\ket{\psi}\ket{\downarrow_z^{pointer}}\ket{q_0}\rightarrow 
e^{-i\hat{H}_0\left(\tau+q_0/v\right)}\Big(\ket{f}\bra{f}\ket{i}\ket{\uparrow_z^{pointer}} 
+\ket{f_\perp}\bra{f_\perp}\ket{i}\ket{\downarrow_z^{pointer}}\Big)\ket{q_0+v\tau}.
\end{equation}
The pre-factor $e^{-i\hat{H}_0\left(\tau+q_0/v\right)}$ simply indicates spin-precession under $\hat{H}_0$ in the time, $\tau+q_0/v$, elapsed after the measurement interaction. Directly measuring if the pointer is in $\ket{\uparrow_z^{pointer}}$ or $\ket{\downarrow_z^{pointer}}$ (note that the WAY theorem has no bearing on this, since the pointer has no Hamiltonian) projects the qubit onto the pure state $e^{-i\hat{H}_0\left(\tau+q_0/v\right)}\ket{f}$ with probability $|\bra{f}\ket{i}|^2$ and $e^{-i\hat{H}_0\left(\tau+q_0/v\right)}\ket{f_\perp}$ with probability $|\bra{f_\perp}\ket{i}|^2$. The clock is separable from the qubit, and the measurement of $\ket{f}\bra{f}$ is ideal, regardless of what pure qubit state $\ket{f}$ is. The measurement is perfect, but energy uncertainty is infinite, which is impractical. Also, with infinite energy uncertainty, there is no meaningful clock energy shift, which is what our article calculates. Instead, in our paper, we consider a clock with large, but finite, energy uncertainty. This maintains qualitative similarity to the ideal case in that the measurement is accurate, while at the same time being more realistic and allowing a meaningful assignment of the clock mean energy shift. Unlike in the ideal case, Eq. (\ref{appendixEq:evIdeal}), the clock is weakly entangled to the qubit-pointer subsystem prior to the pointer measurement, and hence the clock energy depends on the particular post-selection obtained, as our paper illustrates.

At the other extreme, we may consider the clock to have a well-defined energy. If the clock is initialized in an energy eigenstate, $\ket{p}$, then there is no meaningful sense in which the clock \textit{passes the interaction region} (the clock's $q$-distribution is constant in time and not normalizable) and the measurement ``happens.'' This in itself illustrates the WAY theorem, but it is better to consider the clock to have finite, but small energy uncertainty. As an example, we consider the initial clock state to be $\bra{q}\ket{\phi}=(2\pi\sigma_q^2)^{-1/4}e^{-(q-q_0)^2/(4\sigma_q^2)}$, where $q_0<0$, and $\sigma_q\gg\frac{v}{\omega_0}$ is large so that the clock energy uncertainty $\frac{v}{2\sigma_q}$ is small compared to the qubit level spacing, $\omega_0$. The initial mean clock position $|q_0|\gg\sigma_q$ should be larger still so that the clock is meaningfully ``left'' of the interaction region at the start. We take the initial qubit state to be the energy eigenstate, $\ket{\uparrow_z}$, since the mathematics is simplified, while still being illustrative. If we project on the $\ket{\uparrow_z}$ pointer state after the three-body measurement interaction, the evolution is described, for large $\tau$, by:
\begin{equation}\label{appendixEq:evSubOptimal}
\ket{\uparrow_z}\ket{\downarrow_z^{pointer}}\int\ket{q}\bra{q}\ket{\phi}dq\rightarrow
\int e^{-i\hat{H}_0\left(\tau+q/v\right)}\ket{f}\bra{f}e^{+i\hat{H}_0q/v}\ket{\uparrow_z}\ket{q+v\tau}\bra{q}\ket{\phi}dq.
\end{equation}
The above is essentially exact, the only approximation made was to ignore the minimal contributions from position eigenkets that did not cross the interaction region (thus it is applies for small $\sigma_q$ as well). The pointer state, being $\ket{\uparrow_z^{pointer}}$ is dropped in the second line. Tracing out the clock gives the reduced qubit density matrix:
\begin{equation}\label{appendixEq:rhoFRep}
\hat{\rho}_{0}=\int e^{-i\hat{H}_0q/v}\ket{f}\bra{f}e^{+i\hat{H}_0q/v}|\bra{q-v\tau}\ket{\phi}|^2dq.
\end{equation}
In the above integral, each $q/v$ may be interpreted as a time elapsed. The reduced qubit state is equivalent to that if the qubit evolved from $\ket{f}$ for a \textit{statistically} uncertain time $q/v$, with probability density $|\bra{q-v\tau}\ket{\phi}|^2$ for the different possible times. Since the qubit non-trivially precesses under $\hat{H}_0$ (if $\ket{f}$ is not an energy eigenstate) for this statistically uncertain time, the final qubit state is statistically uncertain, i.e. the qubit state is a mixed state (more so the higher the energy uncertainty of $\ket{f}$), as opposed to a pure state that would occur in an ideal measurement. This is better seen by expressing $\hat{\rho}_0$ in terms of energy eigenkets: $\hat{\rho}_0=\rho_{\uparrow\uparrow}\ket{\uparrow}\bra{\uparrow}+\rho_{\uparrow\downarrow}\ket{\uparrow}\bra{\downarrow}+\overline{\rho_{\uparrow\downarrow}}\ket{\downarrow}\bra{\uparrow}+\rho_{\downarrow\downarrow}\ket{\downarrow}\bra{\downarrow}$. In particular, the off-diagonal element $\rho_{\uparrow\downarrow}$ is:
\begin{equation}\label{appendixEq:rhoUpDown}
\begin{split}
\rho_{\uparrow\downarrow}&=\bra{\uparrow_z}\ket{f}\bra{f}\ket{\downarrow_z}
(2\pi\sigma_q^2)^{-1/2}\int_{-\infty}^{\infty} e^{-i\omega_0q/v}e^{-(q-q_0-v\tau)^2/(2\sigma_q^2)}dq\\*
&=\bra{\uparrow_z}\ket{f}\bra{f}\ket{\downarrow_z}e^{-i\omega_0(q_0+v\tau)/v}e^{-\frac{\omega_0^2\sigma_q^2}{2v^2}}
\end{split}
\end{equation}
In the case of well-defined energy, in particular $\sigma_q\gg\frac{v}{\omega_0}$, the off-diagonal element vanishes, so that the qubit state is a statistical mixture of $\ket{\uparrow_z}$ and $\ket{\downarrow_z}$. As $\sigma_q\rightarrow0$, on the other hand, the qubit state approaches a pure superposition of $\ket{\uparrow_z}$ and $\ket{\downarrow_z}$, i.e. the measurement is more ideal.

\subsection{Jaynes-Cummings model}
In the Jaynes-Cummings model, the oscillator plays the role of the ancillary system with large energy (photon number) uncertainty. In this case, the energy uncertainty is necessary in order for the qubit-oscillator coupling $\hat{H}_{JC}=-i\frac{\Omega_0}{2}\left(\hat{a}\hat{\sigma}_+-\hat{a}^\dagger\hat{\sigma}_-\right)$ to generate (approximately) pure qubit rotations. As stated in the introduction, results similar to the WAY theorem apply when the goal is state transformation, as opposed to measurement \cite{popescu2018quantum,popescu2020reference}.

At one extreme, we may consider that the initial oscillator state is a Fock state $\ket{n}$, where $n>0$. If the initial qubit state is $\ket{\uparrow_z}$, then the total number of quanta is well-defined, with value $n+1$. Under the action of the evolution operator $\hat{U}=e^{-i\hat{H}_{JC}t}$, the qubit-oscillator state becomes:
\begin{equation}\label{appendixEq:evOscillatorSubOptimal}
\hat{U}\ket{\uparrow_z}\ket{n}=\cos\frac{\Omega_0t\sqrt{n+1}}{2}\ket{\uparrow_z}\ket{n}
+\sin\frac{\Omega_0t\sqrt{n+1}}{2}\ket{\downarrow_z}\ket{n+1}.
\end{equation}
This is clearly not a pure qubit rotation. The qubit and oscillator are entangled for most $t$ (for example, at $t=\frac{\pi}{2\Omega_0\sqrt{n+1}}$, there is maximal entanglement); their energies are correlated so as to ensure that the number of quanta remains the inital value, $n+1$.

At the other extreme, we may consider, as we do in our paper, that the initial oscillator state is a coherent state $\ket{\alpha}$ (where $\alpha$ is real) with a large photon number (and hence large photon number uncertainty) $n_0=\alpha^2$. Then, under the action of the evolution operator $\hat{U}=e^{-i\hat{H}_{JC}t}$, the qubit-oscillator state satisfies:
\begin{equation}\label{appendixEq:evOscillatorMoreOptimal}
\bra{n}\hat{U}\ket{\uparrow_z}\ket{\alpha}=\bra{n}\ket{\alpha}\cos\frac{\Omega_0t\sqrt{n+1}}{2}\ket{\uparrow_z}
+\bra{n-1}\ket{\alpha}\sin\frac{\Omega_0t\sqrt{n}}{2}\ket{\downarrow_z},
\end{equation}
for $n\geq1$. Furthermore, we take $\Omega_0t\sqrt{n_0}\approx\theta$, i.e. as we consider higher initial mean photon numbers $n_0$, we also take the coupling strength to fall as $1/\sqrt{n_0}$. Then, Eq. (\ref{appendixEq:evOscillatorMoreOptimal}) becomes:
\begin{equation}\label{appendixEq:evOscillatorOptimal}
\bra{n}\hat{U}\ket{\uparrow_z}\ket{\alpha}=\bra{n}\ket{\alpha}\cos\left(\frac{\theta}{2}\sqrt{1+\frac{n-n_0+1}{n_0}}\right)\ket{\uparrow_z}
+\bra{n-1}\ket{\alpha}\sin\left(\frac{\theta}{2}\sqrt{1+\frac{n-n_0+1}{n_0}}\right)\ket{\downarrow_z}.
\end{equation}
Since $\bra{n}\ket{\alpha}$ is small for $n$ not within $\sqrt{n_0}$ of $n_0$, $n-n_0$ is of order $\sqrt{n_0}$ for photon numbers worthy of consideration. $\frac{n-n_0+1}{n_0}$ thus scales like $1/\sqrt{n_0}$ for these $n$, and, as we take the limit as $n_0\rightarrow\infty$, $\bra{n}\hat{U}\ket{\uparrow_z}\ket{\alpha}\sim\bra{n}\ket{\alpha}\left(\cos\frac{\theta}{2}\ket{\uparrow_z}+\sin\frac{\theta}{2}\ket{\downarrow_z}\right)$ (here, we have also used $\bra{n}\ket{\alpha}\sim\bra{n-1}\ket{\alpha}$). The qubit state loses its dependence on $n$, and approaches the result of a pure rotation: $\cos\frac{\theta}{2}\ket{\uparrow_z}+\sin\frac{\theta}{2}\ket{\downarrow_z}=e^{-i\frac{\theta}{2}\hat{\sigma}_y}\ket{\uparrow_z}$.

\section{Quantum Clock Model (Evolution)}

Here we provide details which should be helpful for understanding the derivation of the global system evolution in the quantum clock model (Eq. (\ref{eq:evBefore})-(\ref{eq:evAfter})). We provide a similar model whose state evolution is asymptotic to that of the clock model in the appropriate limit. We derive global state evolution in this model and then explain why the results apply to the quantum clock model as well. We give two methods of solving the problem. Readers not satisfied with the rigor of the first method may prefer the second, scattering method.

\subsection{First method}
The quantum clock model Hamiltonian (Eq. (\ref{eq:clockModelHamiltonian})) is much like the Hamiltonian:
\begin{equation}
\begin{split}
\hat{H}&=\left(\frac{\omega_0}{2}\hat{\sigma}_z+v\hat{p}\right)\hat{\Pi}_{out}
+\left(-\frac{\pi v}{2L}\ket{f}\bra{f}\hat{\sigma}_y^{pointer}+v\hat{p}\right)\hat{\Pi}_{in}\\*
&\equiv\hat{H}_{out}\hat{\Pi}_{out}+\hat{H}_{in}\hat{\Pi}_{in},
\end{split}
\label{appendixEq:appendixHamiltonian}
\end{equation}
where we have implicitly defined $\hat{H}_{out}$ and $\hat{H}_{in}$, and
\begin{subequations}
\begin{equation}
\hat{\Pi}_{in}=\int_{-L/2}^{+L/2} dq \ket{q}\bra{q}
\end{equation}
\begin{equation}
\hat{\Pi}_{out}=1-\hat{\Pi}_{in}.
\end{equation}
\end{subequations}
Let $\ket{\Psi}$ be an arbitrary quantum state of the qubit-pointer subsystem. Clearly, we have $\hat{H}\ket{\Psi}\ket{q}=\hat{H}_{in}\ket{\Psi}\ket{q}$ for $|q|<L/2$ and $\hat{H}\ket{\Psi}\ket{q}=\hat{H}_{out}\ket{\Psi}\ket{q}$ for $|q|>L/2$. It is helpful to see what evolutions $\hat{H}_{in}$ and $\hat{H}_{out}$ would separately generate. Let $\ket{\psi}$ be a quantum state of the qubit, and $\ket{\chi}$ a state of the pointer.

\begin{equation}
\begin{split}
&e^{-it\hat{H}_{in}}\ket{\psi}\ket{\chi}\ket{q}\\*
&=\left(\ket{f}\bra{f}\ket{\psi}e^{+i\frac{\pi vt}{2L}\hat{\sigma}_y^{pointer}}\ket{\chi}+\ket{f_\perp}\bra{f_\perp}\ket{\psi}\ket{\chi}\right)e^{-ivt\hat{p}}\ket{q}\\*
&=\left(\ket{f}\bra{f}\ket{\psi}\hat{R}_y^{pointer}(-\frac{\pi vt}{L})\ket{\chi}+\ket{f_\perp}\bra{f_\perp}\ket{\psi}\ket{\chi}\right)\ket{q+vt}.\\*
\end{split}
\end{equation}
Here we have used the fact that $e^{-i\frac{\theta}{2}\hat{\sigma}_y}=\hat{R}_y(\theta)$ is a rotation operator and $e^{-i\hat{p}\Delta}=\hat{T}(\Delta)$ is a translation operator. The effect of $\hat{H}_{in}$ is to rotate the pointer \textbf{conditional} on the state of the target qubit, while translating the clock at a rate $v$.

For $\hat{H}_{out}$, we have:
\begin{equation}
\begin{split}
e^{-it\hat{H}_{out}}\ket{\psi}\ket{\chi}\ket{q}\\*
=e^{-i\frac{\omega_0 t}{2}\hat{\sigma}_z}\ket{\psi}\ket{\chi}e^{-ivt\hat{p}}\ket{q}\\*
=e^{-i\frac{\omega_0 t}{2}\hat{\sigma}_z}\ket{\psi}\ket{\chi}\ket{q+vt}.
\end{split}
\end{equation}
The effect of $\hat{H}_{out}$ is to rotate the target qubit, \textbf{unconditional} on the state of the pointer, while translating the clock at a rate $v$.

Now suppose we initialize the clock in $\ket{q}$, where $q<-L/2$, and the qubit and pointer are in states $\ket{\psi}$ and $\ket{\chi}$, respectively. Note that, although $|q|>L/2$ and $\hat{H}\ket{\psi}\ket{\chi}\ket{q}=\hat{H}_{out}\ket{\psi}\ket{\chi}\ket{q}$, we do \textbf{not} necessarily have $e^{-it\hat{H}}\ket{\psi}\ket{\chi}\ket{q}=e^{-it\hat{H}_{out}}\ket{\psi}\ket{\chi}\ket{q}$. However, we do have $e^{-it\hat{H}}\ket{\psi}\ket{\chi}\ket{q}=e^{-it\hat{H}_{out}}\ket{\psi}\ket{\chi}\ket{q}$ to first order in $t$, so the evolutions are the same for short times. Consider what happens over a short time $\delta{t}$. $e^{-i\delta{t}\hat{H}}\ket{\psi}\ket{\chi}\ket{q}=e^{-i\delta{t}\hat{H}_{out}}\ket{\psi}\ket{\chi}\ket{q}=e^{-i\frac{\omega_0 \delta{t}}{2}\hat{\sigma}_z}\ket{\psi}\ket{\chi}\ket{q+v\delta{t}}$. Provided that $\delta{t}$ is very small, $q+v\delta{t}<-L/2$ and we know that another short evolution will also be governed by the effective Hamiltonian $\hat{H}_{out}$. By induction, we have that $e^{-it\hat{H}}\ket{\psi}\ket{\chi}\ket{q}=e^{-it\hat{H}_{out}}\ket{\psi}\ket{\chi}\ket{q}$, provided that $t$ is not so large as to cause the clock to arrive at $-L/2$, i.e. $t<(-L/2-q)/v$. By similar logic, we have, for large enough $t$, that:
\begin{equation}
e^{-it\hat{H}}\ket{\psi}\ket{\chi}\ket{q}=e^{-it_3\hat{H}_{out}}e^{-it_2\hat{H}_{in}}e^{-it_1\hat{H}_{out}}\ket{\psi}\ket{\chi}\ket{q},
\end{equation}
where $t_1=(-L/2-q)/v$, $t_2=L/v$, and $t_3=t-t_1-t_2$. First, the system acts as if under the effective Hamiltonian $\hat{H}_{out}$, which causes the clock to arrive at the interaction region. Then, the system acts as if under the effective Hamiltonian $\hat{H}_{in}$, until this causes the clock to exit the interaction region. Lastly, the system acts as if under the effective Hamiltonian $\hat{H}_{out}$ again, since the clock has exited the interaction region, and future evolution will only cause it to move farther away.

There are a few details left to fill in. Since $t_2=L/v$, the pointer's $Y$ rotation is by an angle $-\pi$. Thus, if the pointer is prepared in a state of maximal $\hat{\sigma}_y$ uncertainty (such as $\ket{\downarrow_z^{pointer}}$), this rotation causes it to evolve to an orthogonal state ($\ket{\uparrow_z^{pointer}}$). Any angle $\pi \mod 2\pi$ would work, but the $-\pi$ rotation happens to give a nice $+1$ phase in going from $\ket{\downarrow_z^{pointer}}$ to $\ket{\uparrow_z^{pointer}}$, a transition present in Eq. (\ref{eq:evAcross}) of the main text.

More important are the differences (and similarities) between the Hamiltonians in Eq. (\ref{eq:clockModelHamiltonian}) and (\ref{appendixEq:appendixHamiltonian}). One difference is that $\hat{H}_{in}$ (which occurs in Eq. (\ref{appendixEq:appendixHamiltonian})) does not include a local qubit Hamiltonian, as is present in the clock model Hamiltonian (Eq. (\ref{eq:clockModelHamiltonian})); since we aim to study the qubit energy, the qubit Hamiltonian should always be included. In this section, though, we concerned ourselves only with evolution of the global system state. We note that in the clock model, the interaction region is very narrow (represented by a delta function). Since the local qubit Hamiltonian in the clock model is finite, its contribution to the evolution during the infinitesimal transit time is negligible. The system state evolution in the two models is asymptotic as the interaction region's width goes to zero ($L\rightarrow 0$). This gives the results of Eq. (\ref{eq:evBefore})-(\ref{eq:evAfter}).

\subsection{Scattering method}
Alternatively, to solve for the global system evolution due to the modified Hamilton (Eq. (\ref{appendixEq:appendixHamiltonian})), we could solve for the energy eigenstates and eigenvalues. To understand how the math works out, it is best to start with just understanding the pointer flip, so first we consider the simpler Hamiltonian:

\begin{equation}
\hat{H}_{flip}=v\hat{p}-\frac{\pi{v}}{2L}\hat{\sigma}_y\hat{\Pi}_{in},
\end{equation}
where $\hat{p}=-i\frac{\partial}{\partial{q}}$. As an ansatz, we assume the eigenvectors take the form: $\ket{E_\pm}=\int{dq}\phi_E^{\pm}(q)\ket{q}\ket{\pm_y}$. These eigenvectors are degenerate: $\hat{H}_{flip}\ket{E_\pm}=E\ket{E_\pm}$. By solving the eigenvalue equation in each region and demanding continuity of wavefunctions at the boundaries $\pm L/2$, we find:

\begin{equation}
\begin{split}
&\phi_E^{\pm}(q)\ket{\pm_y} =\\*
&\begin{cases}
          \frac{1}{\sqrt{2\pi{v}}}e^{iqE/v}\ket{\pm_y} & q < -L/2 \\*
          \frac{1}{\sqrt{2\pi{v}}}e^{iq\left(E+\frac{\pi{v}}{2L}\hat{\sigma}_y\right)/v}e^{i\frac{\pi}{4}\hat{\sigma}_y}\ket{\pm_y} & |q| \leq L/2 \\*
          \frac{1}{\sqrt{2\pi{v}}}e^{iqE/v}e^{i\frac{\pi}{2}\hat{\sigma}_y}\ket{\pm_y} & q > L/2. \\*
     \end{cases}
\end{split}
\end{equation}
We see that there is a phase shift across the interaction region, conditional on the $y$-spin.

Since the Hamiltonian is degenerate, we can take superpositions of degenerate eigenvectors to produce eigenvectors with the same eigenvalue. For example, $\ket{E_\chi}=\ket{E_+}\bra{+_y}\ket{\chi}+\ket{E_-}\bra{-_y}\ket{\chi}$, where $\ket{\chi}$ is an arbitrary spinor, satisfies $\hat{H}_{flip}\ket{E_\chi}=E\ket{E_\chi}$. Further, we have:

\begin{equation}
\begin{split}
&\bra{q}\ket{E_\chi} =\\*
&\begin{cases}
          \frac{1}{\sqrt{2\pi{v}}}e^{iqE/v}\ket{\chi} & q < -L/2 \\*
          \frac{1}{\sqrt{2\pi{v}}}e^{iqE/v}\left(e^{i\left(\frac{q+L/2}{L}\right)\frac{\pi}{2}\hat{\sigma}_y}\ket{\chi}\right) & |q| \leq L/2 \\*
          \frac{1}{\sqrt{2\pi{v}}}e^{iqE/v}\left(e^{i\frac{\pi}{2}\hat{\sigma}_y}\ket{\chi}\right) & q > L/2. \\*
     \end{cases}
\end{split}
\label{appendixEq:EChi}
\end{equation}
For $\ket{\chi}\neq\ket{\pm_y}$, the spin of $\ket{E_\chi}$ correlates with position. Left of the interaction region, the spin is fixed at $\ket{\chi}$. In the interaction region, the spin continuously varies (linearly) from $\ket{\chi}$ to $e^{i\frac{\pi}{2}\hat{\sigma}_y}\ket{\chi}$. And right of the interaction region, the spin is fixed at $e^{i\frac{\pi}{2}\hat{\sigma}_y}\ket{\chi}$. For $\ket{\chi}=\ket{-_z}$, $e^{i\frac{\pi}{2}\hat{\sigma}_y}\ket{\chi}=\ket{+_z}$. The ``left'' and ``right'' spinors are orthogonal.

Now we evolve an initial state $\ket{q}\ket{\chi}$, where $q<-L/2$, using linearity of the Schr\"{o}dinger equation. Note that, for all $E$, $\ket{q}\ket{\chi}$ has no projection onto $\ket{E_{\chi_\perp}}=\ket{E_+}\ket{+_y}\bra{+_y}\ket{\chi_\perp}+\ket{E_-}\bra{-_y}\ket{\chi_\perp}$ (where $\bra{\chi_\perp}\ket{\chi}=0$). Thus,

\begin{equation}
\begin{split}
&e^{-it\hat{H}_{flip}}\ket{q}\ket{\chi}=\int dE \ e^{-itE} \ket{E_\chi}\bra{E_\chi} \ \ket{q}\ket{\chi}=\frac{1}{\sqrt{2\pi{v}}}\int dE \ e^{-i\left(\frac{q+vt}{v}\right)E}\ket{E_\chi}\\*
&=\frac{1}{2\pi{v}}\int dE \ e^{-i\left(\frac{q+vt}{v}\right)E}\Bigg[\int_{-\infty}^{-L/2} dq'e^{iq'E/v}\ket{q'}\ket{\chi}+\int_{-L/2}^{L/2} dq'e^{iq'E/v}\ket{q'}\left(e^{i\frac{q'+L/2}{L}\frac{\pi}{2}\hat{\sigma}_y}\ket{\chi}\right)\\*
&+\int_{L/2}^{\infty}dq'e^{iq'E/v}\ket{q'}\left(e^{i\frac{\pi}{2}\hat{\sigma}_y}\ket{\chi}\right)\Bigg]\\*
&=\int_{-\infty}^{-L/2} dq'\delta\left(q'-q-vt\right)\ket{q'}\ket{\chi}+\int_{-L/2}^{L/2} dq'\delta\left(q'-q-vt\right)\ket{q'}\left(e^{i\frac{q'+L/2}{L}\frac{\pi}{2}\hat{\sigma}_y}\ket{\chi}\right)\\*
&+\int_{L/2}^{\infty}dq'\delta\left(q'-q-vt\right)\ket{q'}\left(e^{i\frac{\pi}{2}\hat{\sigma}_y}\ket{\chi}\right).
\end{split}
\label{appendixEq:evolutionDerivationSimple}
\end{equation}

Above, we have changed the order of integration and used the identity $\frac{1}{2\pi}\int dE e^{i\frac{q'-q_0}{v}E}=v\delta(q'-q_0)$. From this equation, it is clear what the state is at each point in time. The clock state is $\ket{q+vt}$, while the qubit state is the corresponding spinor at the clock position (see Eq. (\ref{appendixEq:EChi})).

With this simple example in hand, we move on to solving for the state evolution caused by the Hamiltonian in Eq. (\ref{appendixEq:appendixHamiltonian}). We make ansatz's for the energy eigenstates based on Eq. (\ref{appendixEq:EChi}). Outside of the interaction region, the qubit should rotate, while inside the interaction region, the pointer should rotate. We label our ansatz's as $\ket{E_{f,\chi}}$ and $\ket{E_{\perp,\chi}}$, where the first index labels the qubit state \textbf{at the left boundary} $-L/2$ and the second index $\chi$ labels the pointer state at the left boundary. We are able to satisfy the boundary conditions and eigenvalue equation $\hat{H}=E\ket{E_{f,\chi}}$ with:

\begin{equation}
\begin{split}
&\bra{q}\ket{E_{f,\chi}} =\\*
&\begin{cases}
          \frac{1}{\sqrt{2\pi{v}}}e^{iqE/v}\left(e^{i\frac{\omega_0}{2}\hat{\sigma}_z\left(\frac{-L/2-q}{v}\right)}\ket{f}\right)\ket{\chi} & q < -L/2 \\*
          \frac{1}{\sqrt{2\pi{v}}}e^{iqE/v}\ket{f}\left(e^{i\left(\frac{q+L/2}{L}\right)\frac{\pi}{2}\hat{\sigma}_y}\ket{\chi}\right) & |q| \leq L/2 \\*
          \frac{1}{\sqrt{2\pi{v}}}e^{iqE/v}\left(e^{-i\frac{\omega_0}{2}\hat{\sigma}_z\left(\frac{q-L/2}{v}\right)}\ket{f}\right)\left(e^{i\frac{\pi}{2}\hat{\sigma}_y}\ket{\chi}\right) & q > L/2. \\*
     \end{cases}
\end{split}
\label{appendixEq:EfChi}
\end{equation}
The eigenvalue equation is also satisfied by $\ket{E_\perp,\chi}$, where:
\begin{equation}
\begin{split}
&\bra{q}\ket{E_{\perp,\chi}} =\\*
&\begin{cases}
          \frac{1}{\sqrt{2\pi{v}}}e^{iqE/v}\left(e^{i\frac{\omega_0}{2}\hat{\sigma}_z\left(\frac{-L/2-q}{v}\right)}\ket{f_\perp}\right)\ket{\chi} & q < -L/2 \\*
          \frac{1}{\sqrt{2\pi{v}}}e^{iqE/v}\ket{f_\perp}\ket{\chi} & |q| \leq L/2 \\*
          \frac{1}{\sqrt{2\pi{v}}}e^{iqE/v}\left(e^{-i\frac{\omega_0}{2}\hat{\sigma}_z\left(\frac{q-L/2}{v}\right)}\ket{f_\perp}\right)\ket{\chi} & q > L/2. \\*
     \end{cases}
\end{split}
\label{appendixEq:EperpChi}
\end{equation}

By an analogous derivation to that of Eq. (\ref{appendixEq:evolutionDerivationSimple}), we have, for $q<-L/2$:

\begin{equation}
\begin{split}
&e^{-it\hat{H}}\ket{q}\left(e^{i\frac{\omega_0}{2}\hat{\sigma}_z\left(\frac{-L/2-q}{v}\right)}\ket{f}\right)\ket{\chi}=\\*
&\int_{-\infty}^{-L/2} dq'\delta\left(q'-q-vt\right)\ket{q'}\left(e^{i\frac{\omega_0}{2}\hat{\sigma}_z\left(\frac{-L/2-q'}{v}\right)}\ket{f}\right)\ket{\chi}+\\*
&\int_{-L/2}^{L/2} dq'\delta\left(q'-q-vt\right)\ket{q'}\ket{f}\left(e^{i\left(\frac{q'+L/2}{L}\right)\frac{\pi}{2}\hat{\sigma}_y}\ket{\chi}\right)+\\*
&\int_{L/2}^{\infty}dq'\delta\left(q'-q-vt\right)\ket{q'}\left(e^{-i\frac{\omega_0}{2}\hat{\sigma}_z\left(\frac{q'-L/2}{v}\right)}\ket{f}\right)\left(e^{i\frac{\pi}{2}\hat{\sigma}_y}\ket{\chi}\right),
\end{split}
\label{appendixEq:evolutionDerivationComplicated1}
\end{equation}

and:

\begin{equation}
\begin{split}
&e^{-it\hat{H}}\ket{q}\left(e^{i\frac{\omega_0}{2}\hat{\sigma}_z\left(\frac{-L/2-q}{v}\right)}\ket{f_\perp}\right)\ket{\chi}=\\*
&\int_{-\infty}^{-L/2} dq'\delta\left(q'-q-vt\right)\ket{q'}\left(e^{i\frac{\omega_0}{2}\hat{\sigma}_z\left(\frac{-L/2-q'}{v}\right)}\ket{f_\perp}\right)\ket{\chi}\\*
&+\int_{-L/2}^{L/2} dq'\delta\left(q'-q-vt\right)\ket{q'}\ket{f_\perp}\ket{\chi}+\\*
&\int_{L/2}^{\infty}dq'\delta\left(q'-q-vt\right)\ket{q'}\left(e^{-i\frac{\omega_0}{2}\hat{\sigma}_z\left(\frac{q'-L/2}{v}\right)}\ket{f_\perp}\right)\ket{\chi}.
\end{split}
\label{appendixEq:evolutionDerivationComplicated2}
\end{equation}
Taking linear combinations of Eq. (\ref{appendixEq:evolutionDerivationComplicated1}) and (\ref{appendixEq:evolutionDerivationComplicated2}) gives the evolution for other pure qubit states. Evolution according to the Hamiltonian in the \textit{actual} quantum clock model (Eq. (\ref{eq:clockModelHamiltonian})) is obtained in the limit as $L\rightarrow0$. This gives the results of Eq. (\ref{eq:evBefore})-(\ref{eq:evAfter}).

\section{Quantum Clock Model (Energy Shift)}
Here we provide details on the derivation of the quantum clock energy shift with post-selection, $\Delta\langle{E_M}\rangle_f$ (Eq. (\ref{eq:clockMeanEnergyChange})). We take the basis state evolutions in the clock model (Eq. (\ref{eq:evBefore})-(\ref{eq:evAfter})), which were justified in Appendix B, as our starting point.

Concatenating Eq. (\ref{eq:evBefore})-(\ref{eq:evAfter}), gives:
\begin{equation}\label{appendixEq:evOverall}
\ket{\psi}\ket{\downarrow_z^{pointer}}\ket{q}\rightarrow
e^{-i\hat{H}_0\left(\tau+q/v\right)}\Big(\ket{f}\bra{f}e^{+i\hat{H}_0q/v}\ket{\psi}\ket{\uparrow_z^{pointer}}+\ket{f_\perp}\bra{f_\perp}e^{+i\hat{H}_0q/v}\ket{\psi}\ket{\downarrow_z^{pointer}}\Big)\ket{q+v\tau},
\end{equation}
where $q<0$ and $\tau>-q/v$. This describes a qubit-pointer-clock state starting left of the interaction region and ending past it. The clock state above is a position eigenstate (with infinite energy uncertainty). As mentioned in the main text, we now take the initial clock state to instead be describe by a narrow wavefunction localized left of the interaction region: $\bra{q}\ket{\phi}=(2\pi\sigma_q^2)^{-1/4}e^{-(q-q_0)^2/(4\sigma_q^2)}$, where $q_0<0$ and $\sigma_q\ll|q_0|$. Similar results will hold for similar distributions. Given this, the qubit state when the clock arrives at the interaction region is approximately $\ket{i}\equiv e^{+i\hat{H}_0q_0/v}\ket{\psi}$, since the arrival time is $-q_0/v$ with small uncertainty (owing to the narrowness of the clock wavefunction). By linearity of the Schr\"{o}dinger equation:

\begin{equation}\label{appendixEq:evOverallSuperposition}
\begin{split}
&\int{dq}\ket{\psi}\ket{\downarrow_z^{pointer}}\ket{q}\bra{q}\ket{\phi}\rightarrow\\*
&\approx\int{dq} \ e^{-i\hat{H}_0\left(\tau+q/v\right)}\Big(\ket{f}\bra{f}e^{+i\hat{H}_0q/v}\ket{\psi}\ket{\uparrow_z^{pointer}}+\ket{f_\perp}\bra{f_\perp}e^{+i\hat{H}_0q/v}\ket{\psi}\ket{\downarrow_z^{pointer}}\Big)\ket{q+v\tau}\bra{q}\ket{\phi}\\*
&=\int{d\tilde{q}} \ e^{-i\hat{H}_0\tilde{q}/v}\Big(\ket{f^\tau}\bra{f}e^{+i\hat{H}_0\tilde{q}/v}\ket{i}\ket{\uparrow_z^{pointer}}+\ket{f_\perp^\tau}\bra{f_\perp}e^{+i\hat{H}_0\tilde{q}/v}\ket{i}\ket{\downarrow_z^{pointer}}\Big)\ket{q_0+\tilde{q}+v\tau}\bra{q_0+\tilde{q}}\ket{\phi}\\*
\end{split}
\end{equation}

Here we have made a change of variables to $\tilde{q}=q-q_0$ and introduced the ``evolved'' final states $\ket{f^\tau}=e^{-i\hat{H}_0\left(\tau+q_0/v\right)}\ket{f}$ and $\ket{f_\perp^\tau}=e^{-i\hat{H}_0\left(\tau+q_0/v\right)}\ket{f_\perp}$. The approximation that has been made above is to neglect contributions to the integral from $q\geq0$. This is justified by the fact that $q_0<0$ and $\bra{q}\ket{\phi}$ is narrow (we can always consider smaller $\sigma_q$). Narrowness of $\bra{q}\ket{\phi}$ will justify future approximations as well; for each approximation we make in what follows, there is a $\sigma_q$ that is small enough for the approximation to be valid.

Now we look at post-selection. Post-selecting on the $\ket{f}$ outcome corresponds to taking the projection onto  $\ket{\uparrow_z^{pointer}}$. Then the qubit is approximately in state $\ket{f^\tau}$ at the end. The resulting clock state is thus, approximately:

\begin{equation}\label{appendixEq:clockPostselection}
\begin{split}
&\bra{\uparrow_z^{pointer}}\bra{f^\tau}\int{d\tilde{q}} \ e^{-i\hat{H}_0\tilde{q}/v}\Big(\ket{f^\tau}\bra{f}e^{+i\hat{H}_0\tilde{q}/v}\ket{i}\ket{\uparrow_z^{pointer}}+\ket{f_\perp^\tau}\bra{f_\perp}e^{+i\hat{H}_0\tilde{q}/v}\ket{i}\ket{\downarrow_z^{pointer}}\Big)\ket{q_0+\tilde{q}+v\tau}\bra{q_0+\tilde{q}}\ket{\phi}\\*
&=\int{d\tilde{q}}\underbrace{\bra{f^\tau}e^{-i\hat{H}_0\tilde{q}/v}\ket{f^\tau}}_{\bra{f}e^{-i\hat{H}_0\tilde{q}/v}\ket{f}}\bra{f}e^{+i\hat{H}_0\tilde{q}/v}\ket{i}\ket{q_0+\tilde{q}+v\tau}\bra{q_0+\tilde{q}}\ket{\phi}\\*
&=\int{d\tilde{q}}\left(1-\frac{i\tilde{q}}{v}\bra{f}\hat{H}_0\ket{f}+\mathcal{O}(\tilde{q}^2)\right)\left(\bra{f}\ket{i}+\frac{i\tilde{q}}{v}\bra{f}\hat{H}_0\ket{i}+\mathcal{O}(\tilde{q}^2)\right)\ket{q_0+\tilde{q}+v\tau}\bra{q_0+\tilde{q}}\ket{\phi}\\*
&=\int{d\tilde{q}}\left(\bra{f}\ket{i}+\frac{i\tilde{q}}{v}\left[\bra{f}\hat{H}_0\ket{i}-\bra{f}\ket{i}\bra{f}\hat{H}_0\ket{f}\right]+\mathcal{O}(\tilde{q}^2)\right)\ket{q_0+\tilde{q}+v\tau}\bra{q_0+\tilde{q}}\ket{\phi}\\*
&\approx\int{d\tilde{q}}\bra{f}\ket{i}\exp\Bigg[\frac{i\tilde{q}}{v}\left(-\bra{f}\hat{H}_0\ket{f}+\frac{\bra{f}\hat{H}_0\ket{i}}{\bra{f}\ket{i}}\right)\Bigg]\ket{q_0+\tilde{q}+v\tau}\bra{q_0+\tilde{q}}\ket{\phi}\\*
&=\bra{f}\ket{i}\hat{T}\left(q_0+v\tau\right)\int{d\tilde{q}}\exp\Bigg[\frac{i\tilde{q}}{v}\underbrace{\left(-\bra{f}\hat{H}_0\ket{f}+\frac{\bra{f}\hat{H}_0\ket{i}}{\bra{f}\ket{i}}\right)}_{\Delta\langle{E_M}\rangle_f+i\Im\frac{\bra{f}\hat{H}_0\ket{i}}{\bra{f}\ket{i}}}\Bigg]\ket{\tilde{q}}\bra{q_0+\tilde{q}}\ket{\phi}\equiv\bra{f}\ket{i}\ket{\phi_f}.
\end{split}
\end{equation}

Here, $\hat{T}$ is the translation operator for $\hat{q}$. The approximation made above was to ignore terms in the integrand from $\bra{f}e^{-i\hat{H}_0\tilde{q}/v}\ket{f} \bra{f}e^{+i\hat{H}_0\tilde{q}/v}\ket{i}$ that are of order $q^n$ with $n\geq2$, but keep the terms of order $1$ and $\tilde{q}$. This is justified by the factor $\bra{q_0+\tilde{q}}\ket{\phi}=(2\pi\sigma_q^2)^{-1/4}e^{-\tilde{q}^2/(4\sigma_q^2)}$ in the integrand, which sets the scale, $\sigma_q$, for $\tilde{q}$ values that contribute significantly to the integral. Lower values of $\sigma_q$ may always be chosen so as to justify keeping the terms of order $1$ and $\tilde{q}$ while dropping the $\mathcal{O}(\tilde{q}^2)$ terms, and some choices of $\ket{i}$ and $\ket{f}$, such as those causing anomalous weak values, will require smaller $\sigma_q$ than others (for more on these types of approximations, see Ref. \cite{duck1989sense}).

In the last line of Eq. (\ref{appendixEq:clockPostselection}), $\int{d\tilde{q}}\ket{\tilde{q}}\bra{q_0+\tilde{q}}\ket{\phi}$ represents the initial clock wavefunction in coordinates $\tilde{q}$ such that the center/peak is at $\tilde{q}=0$. $\int{d\tilde{q}}e^{i\tilde{q}\Delta\langle{E_M}\rangle_f/v}\ket{\tilde{q}}\bra{q_0+\tilde{q}}\ket{\phi}$ then represents a clock wavefunction with momentum shifted from the initial by $\Delta{p}=\Delta\langle{E_M}\rangle_f/v$. The presence of the imaginary part of the weak value (last line) does not affect this energy shift (see Ref. \cite{aharonov1988result} and \cite{dressel2012significance} for more on the imaginary part). The translation operator outside the integral has no influence on the clock energy. Since the clock Hamiltonian is $\hat{H}_{clock}=v\hat{p}$, we have our result, Eq. (\ref{eq:clockMeanEnergyChange}).

\section{Jaynes-Cummings Model (Energy Shift)}
Here we justify the results for the mean energy shift of the oscillator in the Jaynes-Cummings measurement model. As discussed in the main text, this shift is composed of two terms (see Eq. (\ref{eq:MeanEnergyChangeCommonForm}) and (\ref{eq:MeanEnergyChangeCommonForm2}) and their surrounding text).

The second term is equal and opposite to the mean excitation number change of the qubit when it is driven from an energy eigenstate (the result of the $\hat{\sigma}_z$ measurement) to its final state. This requires little justification. The oscillator and qubit are coupled during this period via the Jaynes-Cummings Hamiltonian (Eq. (\ref{eq:JaynesCummingsHamiltonian})), which preserves the total excitation number of qubit and oscillator, and there is only unitary evolution under this Hamiltonian during this period. Thus, any energy change by the qubit during this period is compensated exactly by the oscillator (the qubit and oscillator are assumed to be perfectly on resonance). The mean energy change of the qubit is calculated under the assumption that the final state ($\ket{f}$ if starting in $\ket{e}\equiv\ket{\uparrow_z}$ or $\ket{f_\perp}$ if starting in $\ket{g}\equiv\ket{\downarrow_z}$) targeted by the rotation is accurately achieved; this means going to the semi-classical limit whereby the oscillator satisfies $n_0=\bra{\alpha}\hat{a}^\dagger\hat{a}\ket{\alpha}=\alpha^2\rightarrow\infty$ while the product of the vacuum Rabi frequency and interaction time $\Omega_0t\rightarrow{0}$ as $1/\sqrt{n_0}$ \cite{haroche2006exploring} (the need for such a limit is in keeping with the WAY theorem). Thus the mean photon number change of the oscillator during this period is either $-\left(\bra{f}\ket{\uparrow_z}\bra{\uparrow_z}\ket{f}-1\right)$ or $-\bra{f_\perp}\ket{\uparrow_z}\bra{\uparrow_z}\ket{f_\perp}$, depending on the result ($\uparrow$, $\downarrow$ respectively) of the $\hat{\sigma}_z$ measurement.

The rest of the oscillator energy change is due to the first drive plus the $\hat{\sigma}_z$ measurement being performed and yielding a particular outcome (e.g. $\ket{\uparrow_z}$). This is more complicated to calculate because of the post-selection involved. We will assume that, during the first drive, the coupling strength $\Omega(t)$ is constant at some vacuum Rabi frequency $\Omega_0$ for a period $t$. This gives the evolution operator $\hat{U}_{JC}=e^{-\frac{\Omega_0t}{2}\left(\hat{a}\hat{\sigma}_+-\hat{a}^\dagger\hat{\sigma}_-\right)}$ for the qubit-oscillator system (recall we work in the interaction picture of the resonant Jaynes-Cummings model). Given initial qubit state $\ket{i}$ and initial oscillator state $\ket{\alpha}$ ($\alpha$ real and positive), the resulting state of the oscillator when we run the interaction and post-select on $\ket{\uparrow_z}$ is $\bra{\uparrow_z}\hat{U}_{JC}\ket{i}\ket{\alpha}$, up to a normalization factor. Similarly, if we post-select instead on $\ket{\downarrow_z}$, the resulting oscillator state is $\bra{\downarrow_z}\hat{U}_{JC}\ket{i}\ket{\alpha}$, up to a normalization factor. These states are given by:

\begin{subequations}
\label{appendixEq:alphaChanged} 
\begin{eqnarray}
\bra{\uparrow_z}\hat{U}_{JC}\ket{i}\ket{\alpha}&=&\sum_n \left(\cos\frac{\Omega_0t\sqrt{n+1}}{2}\bra{\uparrow_z}\ket{i}\bra{n}\ket{\alpha} - \sin\frac{\Omega_0t\sqrt{n+1}}{2}\bra{\downarrow_z}\ket{i}\bra{n+1}\ket{\alpha}\right)\ket{n} \label{alphaUp}
\\*
\bra{\downarrow_z}\hat{U}_{JC}\ket{i}\ket{\alpha}&=& \bra{\downarrow_z}\ket{i}\bra{0}\ket{\alpha}\ket{0} + \sum_{n=1}^{\infty} \left( \sin\frac{\Omega_0t\sqrt{n}}{2}\bra{\uparrow_z}\ket{i}\bra{n-1}\ket{\alpha}+\cos\frac{\Omega_0t\sqrt{n}}{2}\bra{\downarrow_z}\ket{i}\bra{n}\ket{\alpha}\right)\ket{n}. \label{alphaDown}
\end{eqnarray}
\end{subequations}

Since the aim is to approximate the $Y$ rotation of the qubit by an angle $-\theta$, we assume $\Omega_0t\sqrt{n_0}\approx-\theta$. More concretely, we will take $\Omega_0t$ to satisfy $\Omega_0t\sqrt{n_0+m}=-\theta$, where $m\ll{\sqrt{n_0}}$. We leave $m$ a variable because the best possible value of $m$ is not immediately obvious, and we want to show that the leading order photon number shift is essentially independent of the choice. Rewriting Eq. (\ref{appendixEq:alphaChanged}), we get:

\small

\begin{subequations}
\label{appendixEq:alphaChangedRewrite} 
\begin{eqnarray}
\bra{\uparrow_z}\hat{U}_{JC}\ket{i}\ket{\alpha}&=&\sum_n \left(\cos\left(\frac{\theta}{2}\sqrt{1+\frac{n-n_0+1-m}{n_0+m}}\right)\bra{\uparrow_z}\ket{i}\bra{n}\ket{\alpha} + \sin\left(\frac{\theta}{2}\sqrt{1+\frac{n-n_0+1-m}{n_0+m}}\right)\bra{\downarrow_z}\ket{i}\bra{n+1}\ket{\alpha}\right)\ket{n}\nonumber\\*
\label{alphaUpRewrite}\\*
\bra{\downarrow_z}\hat{U}_{JC}\ket{i}\ket{\alpha}&=& \sum_{n=1}^{\infty} \left( -\sin\left(\frac{\theta}{2}\sqrt{1+\frac{n-n_0-m}{n_0+m}}\right)\bra{\uparrow_z}\ket{i}\bra{n-1}\ket{\alpha}+\cos\left(\frac{\theta}{2}\sqrt{1+\frac{n-n_0-m}{n_0+m}}\right)\bra{\downarrow_z}\ket{i}\bra{n}\ket{\alpha}\right)\ket{n} \label{alphaDownRewrite}\\*
&+& \bra{\downarrow_z}\ket{i}\bra{0}\ket{\alpha}\ket{0}.\nonumber
\end{eqnarray}
\end{subequations}

\normalsize
Eq. (\ref{appendixEq:alphaChangedRewrite}) is exact. The un-normalized Born rule probabilities for a given photon number (conditioned on the $\hat{\sigma}_z$ measurement outcome) are $P\left( n \;\middle\vert\; \uparrow_z \right)=|\bra{\uparrow_z}\bra{n}\hat{U}_{JC}\ket{i}\ket{\alpha}|^2$ and $P\left( n \;\middle\vert\; \downarrow_z \right)=|\bra{\downarrow_z}\bra{n}\hat{U}_{JC}\ket{i}\ket{\alpha}|^2$. The resulting mean photon numbers are then given exactly by $\langle{n}\rangle_{i\circlearrowright\uparrow}=\frac{\sum_n nP\left( n \;\middle\vert\; \uparrow_z \right)}{\sum_n P\left( n \;\middle\vert\; \uparrow_z \right)}=n_0+\frac{\sum_n (n-n_0)P\left( n \;\middle\vert\; \uparrow_z \right)}{\sum_n P\left( n \;\middle\vert\; \uparrow_z \right)}$ and $\langle{n}\rangle_{i\circlearrowright\downarrow}=\frac{\sum_n nP\left( n \;\middle\vert\; \downarrow_z \right)}{\sum P\left( n \;\middle\vert\; \downarrow_z \right)}=n_0+\frac{\sum_n (n-n_0)P\left( n \;\middle\vert\; \downarrow_z \right)}{\sum P\left( n \;\middle\vert\; \downarrow_z \right)}$. These summations, while discrete, are well approximated by integrals in the large $n_0$ limit, wherein $\bra{n}\ket{\alpha}$ passes to (the square-root of) a continuous Gaussian $\bra{n}\ket{\alpha}\approx(2\pi n_0)^{-1/4}e^{-\frac{1}{4n_0}(n-n_0)^2}$, which suppresses contributions to the integrals from $n$ not within $\sqrt{n_0}$ of $n_0$. This allows us to replace the trigonometric functions of $n$ with their first-order Taylor expansions, and expand the integration region to $\pm\infty$ while negligibly changing the value (in the $n_0\rightarrow\infty$ limit) of the integrals.

For example,

\begin{equation}
\begin{split}
&\langle{n}\rangle_{\uparrow\circlearrowright\uparrow}-n_0\\*
&\approx\frac{(2\pi n_0)^{-1/2}\int_{-\infty}^\infty{dn}\cos^2\left(\frac{\theta}{2}\sqrt{1+\frac{n-n_0+1-m}{n_0+m}}\right)(n-n_0)e^{-\frac{1}{2n_0}(n-n_0)^2}}
{(2\pi n_0)^{-1/2}\int_{-\infty}^\infty{dn}\cos^2\left(\frac{\theta}{2}\sqrt{1+\frac{n-n_0+1-m}{n_0+m}}\right)e^{-\frac{1}{2n_0}(n-n_0)^2}}\\*
&\approx\frac{\cos^2\frac{\theta}{2}(2\pi n_0)^{-1/2}\int_{-\infty}^\infty{dn}\left(1-\frac{[(n-n_0)+1-m]}{n_0+m}\frac{\theta}{2}\tan\frac{\theta}{2}+\mathcal{O}\left(\frac{(n-n_0)^2}{n_0^2}\right)\right)(n-n_0)e^{-\frac{1}{2n_0}(n-n_0)^2}}
{\cos^2\frac{\theta}{2}(2\pi n_0)^{-1/2}\int_{-\infty}^\infty{dn}\left(1-\frac{[(n-n_0)+1-m]}{n_0+m}\frac{\theta}{2}\tan\frac{\theta}{2}+\mathcal{O}\left(\frac{(n-n_0)^2}{n_0^2}\right)\right)e^{-\frac{1}{2n_0}(n-n_0)^2}}\\*
&=\frac{\cos^2\frac{\theta}{2}\left(-\frac{n_0}{n_0+m}\frac{\theta}{2}\tan\frac{\theta}{2}+\mathcal{O}\left(\frac{1}{\sqrt{n_0}}\right)\right)}
{\cos^2\frac{\theta}{2}\left(1+\mathcal{O}\left(\frac{1}{\sqrt{n_0}}\right)\right)}\sim -\frac{\theta}{2}\tan\frac{\theta}{2} \quad \mathrm{as} \quad n_0\rightarrow\infty\\*
&\equiv\Delta\langle{\tilde{E}_M}\rangle_{\uparrow\circlearrowright\uparrow}(\theta).
\end{split}
\end{equation}

In the fourth line above, we have allowed for corrections of order $\frac{1}{\sqrt{n_0}}$ or higher (e.g. $\frac{1}{n_0}$), our point being that while such corrections may exist (especially when replacing the trigonometric functions with higher-order Taylor expansions), these corrections vanish as $n_0\rightarrow\infty$.

As discussed in the main text, $\Delta\langle{\tilde{E}_M}\rangle_{\uparrow\circlearrowright\uparrow}(\theta)$ is the photon number shift when we prepare the qubit in $\ket{\uparrow_z}$, couple to the oscillator via the Jaynes-Cummings interaction so as to (albeit, imperfectly) execute the qubit $Y$ rotation by $-\theta$, measure $\hat{\sigma}_z$, and get the outcome $\ket{\uparrow_z}$. This should be evident from the derivation now. Following the same approximation methods as before, we can calculate the values of $\Delta\langle{\tilde{E}_M}\rangle_{\downarrow\circlearrowright\uparrow}(\theta)$, $\Delta\langle{\tilde{E}_M}\rangle_{\uparrow\circlearrowright\downarrow}(\theta)$, and $\Delta\langle{\tilde{E}_M}\rangle_{\downarrow\circlearrowright\downarrow}(\theta)$ (see Eq. (\ref{eq:oscillatorSubChanges}) of the main text). In these calculations it is notable that $|\bra{n+1}\ket{\alpha}|^2$ may be substituted by a Gaussian centered at $n_0-1$ with variance $n_0$, and $|\bra{n-1}\ket{\alpha}|^2$ may be substituted by a Gaussian centered at $n_0+1$ with variance $n_0$.

Lastly, we calculate the full form (see ``first term'' of Eq. (\ref{eq:MeanEnergyChangeCommonForm})), where $\ket{i}$ may be a superposition of $\ket{\uparrow_z}$ and $\ket{\downarrow_z}$. In these calculations it is notable that $\bra{n}\ket{\alpha}\bra{\alpha}\ket{n+1}=\overline{\bra{n}\ket{\alpha}\bra{\alpha}\ket{n+1}}$ (since $\alpha$ is assumed real) may be substituted by a Gaussian centered at $n_0-\frac{1}{2}$ with variance $n_0$ (up to a negligible normalization factor $e^{-\frac{1}{8n_0}}=1+\mathcal{O}\left(\frac{1}{n_0}\right)$). We find, following the same approximation methods as before, that:

\small

\begin{equation}
\begin{split}
&\langle{n}\rangle_{i\circlearrowright\uparrow}-n_0\sim\\*
&\frac{\cos^2\frac{\theta}{2}|\bra{\uparrow_z}\ket{i}|^2\left(-\frac{\theta}{2}\tan\frac{\theta}{2}\right)+\sin^2\frac{\theta}{2}|\bra{\downarrow_z}\ket{i}|^2\left(-1+\frac{\theta}{2}\cot\frac{\theta}{2}\right)+\sin\frac{\theta}{2}\cos\frac{\theta}{2}\Re\left(\bra{\uparrow_z}\ket{i}\overline{\bra{\downarrow_z}\ket{i}}\right)\left(-\frac{\theta}{2}\tan\frac{\theta}{2}-1+\frac{\theta}{2}\cot\frac{\theta}{2}\right)}
{\cos^2\frac{\theta}{2}|\bra{\uparrow_z}\ket{i}|^2+\sin^2\frac{\theta}{2}|\bra{\downarrow_z}\ket{i}|^2+2\sin\frac{\theta}{2}\cos\frac{\theta}{2}\Re\left(\bra{\uparrow_z}\ket{i}\overline{\bra{\downarrow_z}\ket{i}}\right)}\\*
&=\frac{|\bra{f}\ket{\uparrow_z}\bra{\uparrow_z}\ket{i}|^2\Delta\langle{\tilde{E}_M}\rangle_{\uparrow\circlearrowright\uparrow}(\theta)+|\bra{f}\ket{\downarrow_z}\bra{\downarrow_z}\ket{i}|^2\Delta\langle{\tilde{E}_M}\rangle_{\downarrow\circlearrowright\uparrow}(\theta)+2\Re\left(\bra{f}\ket{\uparrow_z}\bra{\uparrow_z}\ket{i}\overline{\bra{f}\ket{\downarrow_z}\bra{\downarrow_z}\ket{i}}\right)\frac{\Delta\langle{\tilde{E}_M}\rangle_{\uparrow\circlearrowright\uparrow}(\theta)+\Delta\langle{\tilde{E}_M}\rangle_{\downarrow\circlearrowright\uparrow}(\theta)}{2}}
{|\bra{f}\ket{i}|^2},
\end{split}
\end{equation}

\normalsize
as $n_0\rightarrow\infty$. Here, we have made use of the fact that $\ket{f}=\cos\frac{\theta}{2}\ket{\uparrow_z}+\sin\frac{\theta}{2}\ket{\downarrow_z}$. This justifies (the first term of) Eq. (\ref{eq:MeanEnergyChangeCommonForm}). The justification of Eq. (\ref{eq:MeanEnergyChangeCommonForm2}) is analogous and makes use of the equality $\ket{f_\perp}=-\sin\frac{\theta}{2}\ket{\uparrow_z}+\cos\frac{\theta}{2}\ket{\downarrow_z}$.

The magnitude of $n_0$ required for these analytical results (obtained in the $n_0\rightarrow\infty$ limit) to be accurate depends on the specific measurement context (in particular the initial state $\ket{i}$ and rotation angle $-\theta$). Generally speaking, larger shifts require larger $n_0$ (larger photon number variance $n_0$). Fig. A1 demonstrates convergence to the analytical value as $n_0\rightarrow\infty$.

\begin{figure}
    \begin{center}
    \includegraphics[width=0.7\linewidth]{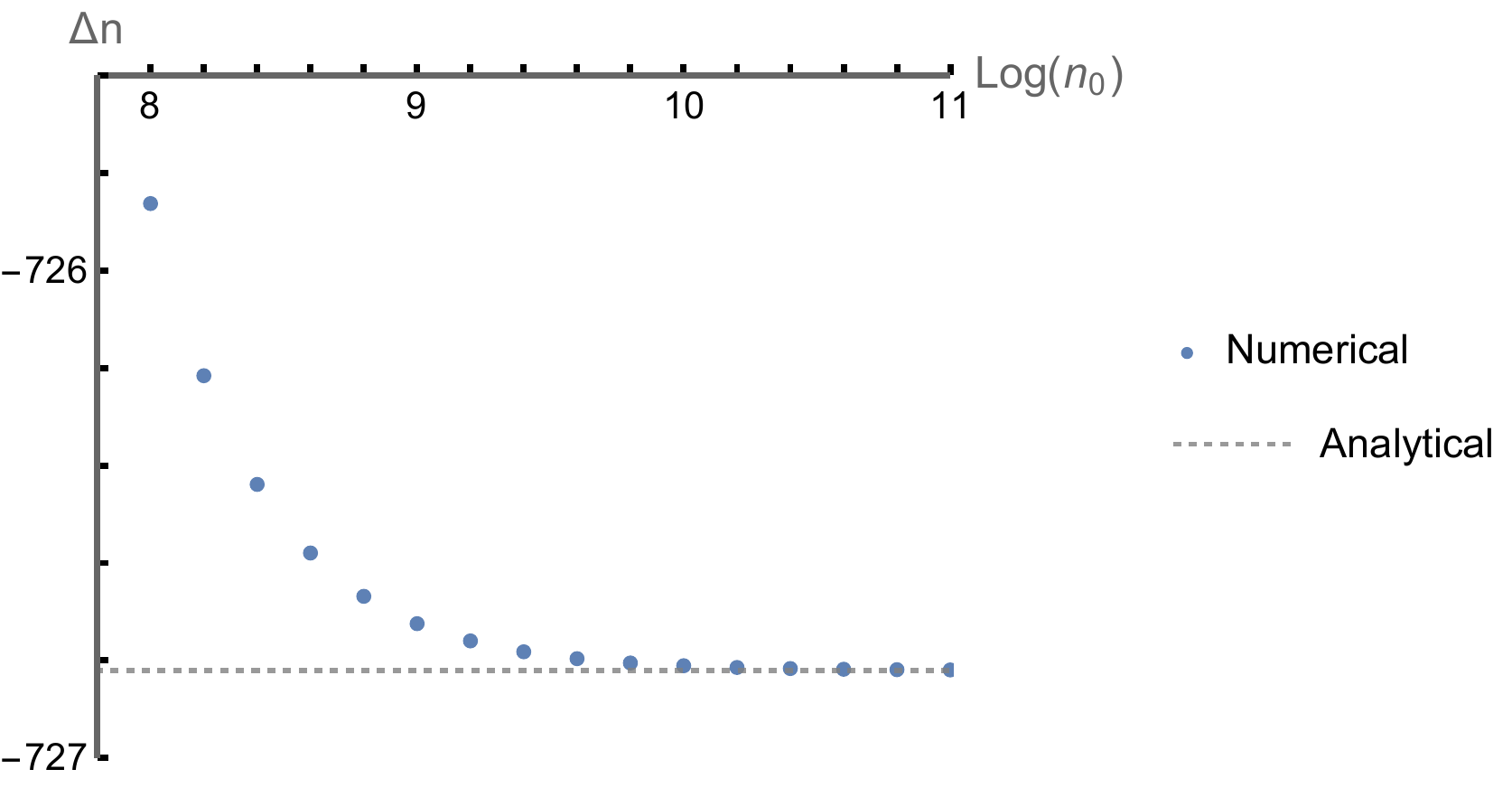}\caption{Photon number shift (denoted $\Delta n$ in the figure) corresponding to the first drive followed by the $\hat{\sigma}_z$ measurement yielding $\ket{\uparrow_z}.$ An extreme case is considered: $\ket{i}=\frac{1}{\sqrt{2}}\left(\ket{\uparrow_z}+\ket{\downarrow_z}\right)$ while $\theta=\pi\left(\frac{3}{2}-\frac{1}{400}\right)$; the first drive is by $-\theta$ and the qubit state after the \textit{second} drive would be $\ket{f}=\cos\frac{\theta}{2}\ket{\uparrow_z}+\sin\frac{\theta}{2}\ket{\downarrow_z}$, which is almost orthogonal to $\ket{i}$. Thus the measurement outcome is very unlikely and we can expect a large shift in the mean energy of the measurement apparatus, similar to an anomalous weak value. Likewise, a large value of the photon number variance $n_0$ is needed to approach the analytical value presented in the main text. Note that $n_0$ is also the initial mean photon number, because a coherent state is used. Numerical values of the photon number shift are calculated using the un-normalized photon number probabilities $P\left( n \;\middle\vert\; \uparrow_z \right)$ from mod-squaring the amplitudes in Eq. (\ref{alphaUpRewrite}) with $m=1$; the full form of the trigonometric functions is used (as opposed to a first-order Taylor expansion) and a continuous Gaussian is substituted for $\bra{n}\ket{\alpha}$. Numerical integrals from $n_0-10\sqrt{n_0}$ to $n_0+10\sqrt{n_0}$ are performed, giving an approximation of $\frac{\sum_n (n-n_0)P\left( n \;\middle\vert\; \uparrow_z \right)}{\sum_n P\left( n \;\middle\vert\; \uparrow_z \right)}$. Comparison with the asymptotic value as $n_0\rightarrow\infty$ (the ``analytical'' value) is given in the plot to show how convergence is achieved as $n_0\rightarrow\infty$.  The logarithmic scale used for the x-axis is base-10. }\label{appendixFig:convergence}
    \end{center}
\end{figure}

\section{Degenerate Qubit-Oscillator Interaction}
An alternative qubit-oscillator interaction, $\hat{H}_D=-i\frac{\Omega}{2}\left(\hat{L}\hat{\sigma}_+-\hat{L}^\dagger\hat{\sigma}_-\right)$, where $\hat{L}^\dagger=\sum_{n=1}^\infty \ket{n}\bra{n-1}=\\ \sum_{n=0}^\infty \ket{n+1}\bra{n}$, has the same eigenvectors as $\hat{H}_{JC}=-i\frac{\Omega}{2}\left(\hat{a}\hat{\sigma}_+-\hat{a}^\dagger\hat{\sigma}_-\right)$, but degenerate eigenvalues $\pm\frac{\Omega}{2}$. In theory, we could implement the same measurement protocol (rotate about $Y$, measure $\hat{\sigma}_z$, then perform the opposite rotation) using a coherent state and this Hamiltonian instead of the Jaynes-Cummings one. Doing so returns the clock result, as we will now show.

We will assume that, during the first drive, the coupling strength $\Omega$ is constant at some vacuum Rabi frequency $\Omega_0$ for a period $t$ such that $\Omega_0t=-\theta$. This gives the evolution operator $\hat{U}_D=e^{-\frac{\theta}{2}\left(\hat{L}\hat{\sigma}_+-\hat{L}^{\dagger}\hat{\sigma}_-\right)}$. Given initial qubit state $\ket{i}$ and initial oscillator state $\ket{\alpha}$, the resulting state of the oscillator when we run the interaction and post-select on $\ket{\uparrow_z}$ is: $\bra{\uparrow_z}\hat{U}_{D}\ket{i}\ket{\alpha}$, up to a normalization factor. Similarly, if we post-select instead on $\ket{\downarrow_z}$, the resulting oscillator state is $\bra{\downarrow_z}\hat{U}_{D}\ket{i}\ket{\alpha}$, up to a normalization factor. These states are given by:

\begin{subequations}
\label{appendixEq:alphaChangedDegenerate} 
\begin{eqnarray}
\bra{\uparrow_z}\hat{U}_{D}\ket{i}\ket{\alpha}&=&\sum_n \left(\cos\frac{\theta}{2}\bra{\uparrow_z}\ket{i}\bra{n}\ket{\alpha} + \sin\frac{\theta}{2}\bra{\downarrow_z}\ket{i}\bra{n+1}\ket{\alpha}\right)\ket{n} \nonumber
\\*
&=&\sum_n \left(\bra{f}\ket{\uparrow_z}\bra{\uparrow_z}\ket{i}\bra{n}\ket{\alpha} + \bra{f}\ket{\downarrow_z}\bra{\downarrow_z}\ket{i}\bra{n+1}\ket{\alpha}\right)\ket{n}
\\*
\bra{\downarrow_z}\hat{U}_{D}\ket{i}\ket{\alpha}&=& \bra{\downarrow_z}\ket{i}\bra{0}\ket{\alpha}\ket{0} + \sum_{n=1}^{\infty} \left( -\sin\frac{\theta}{2}\bra{\uparrow_z}\ket{i}\bra{n-1}\ket{\alpha}+\cos\frac{\theta}{2}\bra{\downarrow_z}\ket{i}\bra{n}\ket{\alpha}\right)\ket{n} \nonumber\\*
&=&\bra{\downarrow_z}\ket{i}\bra{0}\ket{\alpha}\ket{0} + \sum_{n=1}^{\infty} \left( \bra{f_\perp}\ket{\uparrow_z}\bra{\uparrow_z}\ket{i}\bra{n-1}\ket{\alpha}+\bra{f_\perp}\ket{\downarrow_z}\bra{\downarrow_z}\ket{i}\bra{n}\ket{\alpha}\right)\ket{n}.
\end{eqnarray}
\end{subequations}

The un-normalized Born rule probabilities for a given photon number (conditioned on the $\hat{\sigma}_z$ measurement outcome) are $P\left( n \;\middle\vert\; \uparrow_z \right)=|\bra{\uparrow_z}\bra{n}\hat{U}_{D}\ket{i}\ket{\alpha}|^2$ and $P\left( n \;\middle\vert\; \downarrow_z \right)=|\bra{\downarrow_z}\bra{n}\hat{U}_{D}\ket{i}\ket{\alpha}|^2$. The resulting mean photon numbers are then given exactly by $\langle{n}\rangle_{i\circlearrowright\uparrow}=\frac{\sum_n nP\left( n \;\middle\vert\; \uparrow_z \right)}{\sum_n P\left( n \;\middle\vert\; \uparrow_z \right)}=n_0+\frac{\sum_n (n-n_0)P\left( n \;\middle\vert\; \uparrow_z \right)}{\sum_n P\left( n \;\middle\vert\; \uparrow_z \right)}$ and $\langle{n}\rangle_{i\circlearrowright\downarrow}=\frac{\sum_n nP\left( n \;\middle\vert\; \downarrow_z \right)}{\sum P\left( n \;\middle\vert\; \downarrow_z \right)}=n_0+\frac{\sum_n (n-n_0)P\left( n \;\middle\vert\; \downarrow_z \right)}{\sum P\left( n \;\middle\vert\; \downarrow_z \right)}$. These summations, while discrete, are well approximated by integrals in the large $n_0$ limit, wherein $\bra{n}\ket{\alpha}$ passes to (the square-root of) a continuous Gaussian $|\bra{n}\ket{\alpha}|^2\approx(2\pi n_0)^{-1/2}e^{-\frac{1}{2n_0}(n-n_0)^2}$, and the integration region can be expanded to $\pm\infty$ while negligibly changing the value. We find:

\footnotesize

\begin{equation}
\begin{split}
&\langle{n}\rangle_{i\circlearrowright\uparrow}\approx\\*
&\frac{\int dn\left(|\bra{f}\ket{\uparrow_z}\bra{\uparrow_z}\ket{i}|^2e^{-\frac{1}{2n_0}(n-n_0)^2}+|\bra{f}\ket{\downarrow_z}\bra{\downarrow_z}\ket{i}|^2e^{-\frac{1}{2n_0}(n-n_0+1)^2}+e^{-\frac{1}{8n_0}}2\Re\left(\bra{f}\ket{\uparrow_z}\bra{\uparrow_z}\ket{i}\overline{\bra{f}\ket{\downarrow_z}\bra{\downarrow_z}\ket{i}}\right)e^{-\frac{1}{2n_0}(n-n_0+\frac{1}{2})^2}\right)n}
{\int dn\left(|\bra{f}\ket{\uparrow_z}\bra{\uparrow_z}\ket{i}|^2e^{-\frac{1}{2n_0}(n-n_0)^2}+|\bra{f}\ket{\downarrow_z}\bra{\downarrow_z}\ket{i}|^2e^{-\frac{1}{2n_0}(n-n_0+1)^2}+e^{-\frac{1}{8n_0}}2\Re\left(\bra{f}\ket{\uparrow_z}\bra{\uparrow_z}\ket{i}\overline{\bra{f}\ket{\downarrow_z}\bra{\downarrow_z}\ket{i}}\right)e^{-\frac{1}{2n_0}(n-n_0+\frac{1}{2})^2}\right)}\\*
&=n_0+\frac{|\bra{f}\ket{\uparrow_z}\bra{\uparrow_z}\ket{i}|^2(0)+|\bra{f}\ket{\downarrow_z}\bra{\downarrow_z}\ket{i}|^2(-1)+e^{-\frac{1}{8n_0}}2\Re\left(\bra{f}\ket{\uparrow_z}\bra{\uparrow_z}\ket{i}\overline{\bra{f}\ket{\downarrow_z}\bra{\downarrow_z}\ket{i}}\right)(-\frac{1}{2})}
{|\bra{f}\ket{\uparrow_z}\bra{\uparrow_z}\ket{i}|^2+|\bra{f}\ket{\downarrow_z}\bra{\downarrow_z}\ket{i}|^2+e^{-\frac{1}{8n_0}}2\Re\left(\bra{f}\ket{\uparrow_z}\bra{\uparrow_z}\ket{i}\overline{\bra{f}\ket{\downarrow_z}\bra{\downarrow_z}\ket{i}}\right)}\\*
&\approx n_0+\frac{|\bra{f}\ket{\uparrow_z}\bra{\uparrow_z}\ket{i}|^2(0)+|\bra{f}\ket{\downarrow_z}\bra{\downarrow_z}\ket{i}|^2(-1)+2\Re\left(\bra{f}\ket{\uparrow_z}\bra{\uparrow_z}\ket{i}\overline{\bra{f}\ket{\downarrow_z}\bra{\downarrow_z}\ket{i}}\right)(-\frac{1}{2})}
{|\bra{f}\ket{\uparrow_z}\bra{\uparrow_z}\ket{i}|^2+|\bra{f}\ket{\downarrow_z}\bra{\downarrow_z}\ket{i}|^2+2\Re\left(\bra{f}\ket{\uparrow_z}\bra{\uparrow_z}\ket{i}\overline{\bra{f}\ket{\downarrow_z}\bra{\downarrow_z}\ket{i}}\right)}\\*
&=n_0+\frac{|\bra{f}\ket{\uparrow_z}\bra{\uparrow_z}\ket{i}|^2\Delta\langle{\tilde{E}_M}\rangle_{\uparrow\circlearrowright\uparrow}+|\bra{f}\ket{\downarrow_z}\bra{\downarrow_z}\ket{i}|^2\Delta\langle{\tilde{E}_M}\rangle_{\downarrow\circlearrowright\uparrow}+2\Re\left(\bra{f}\ket{\uparrow_z}\bra{\uparrow_z}\ket{i}\overline{\bra{f}\ket{\downarrow_z}\bra{\downarrow_z}\ket{i}}\right)\frac{\Delta\langle{\tilde{E}_M}\rangle_{\uparrow\circlearrowright\uparrow}+\Delta\langle{\tilde{E}_M}\rangle_{\downarrow\circlearrowright\uparrow}}{2}}{|\bra{f}\ket{i}|^2}
\label{appendixEq:degenerateBigDerivation}
\end{split}
\end{equation}

\normalsize
in the limit as $n_0\rightarrow\infty$. This gives the photon number change due to the first drive and $\hat{\sigma}_z$ measurement (conditioned on the outcome $\ket{\uparrow_z}$). Next, there is another photon number change due to the second drive. This shift in the mean photon number is simply equal and opposite to the change in the qubit excitation number in going from $\ket{\uparrow_z}$ to $\ket{f}=\cos\frac{\theta}{2}\ket{\uparrow_z}+\sin\frac{\theta}{2}\ket{\downarrow_z}$ because the interaction preserves the total excitation number of qubit and oscillator. Thus this second shift term is $-\left(\bra{f}\ket{\uparrow_z}\bra{\uparrow_z}\ket{f}-1\right)$. This gives Eq. (\ref{eq:MeanEnergyChangeCommonForm}).

$\Delta\langle{\tilde{E}_M}\rangle_{\uparrow\circlearrowright\uparrow}$ is the photon number shift when we prepare the qubit in $\ket{\uparrow_z}$, couple to the oscillator via the degenerate interaction so as to (albeit, imperfectly) execute the qubit $Y$ rotation by $-\theta$, measure $\hat{\sigma}_z$, and get the outcome $\ket{\uparrow_z}$. This should be evident from the derivation. Similarly, $\Delta\langle{\tilde{E}_M}\rangle_{\downarrow\circlearrowright\uparrow}$ is the photon number shift when we prepare the qubit in $\ket{\downarrow_z}$, couple to the oscillator via the degenerate interaction so as to execute the qubit $Y$ rotation by $-\theta$, measure $\hat{\sigma}_z$, and get the outcome $\ket{\uparrow_z}$. Notably, these are not functions of $\theta$; this is unlike the result obtained using the non-degenerate Jaynes-Cummings interaction.

The derivations of $\Delta\langle{\tilde{E}_M}\rangle_{\uparrow\circlearrowright\downarrow}$, $\Delta\langle{\tilde{E}_M}\rangle_{\downarrow\circlearrowright\downarrow}$, and Eq. (\ref{eq:MeanEnergyChangeCommonForm2}) are analogous. In the next section, we show that (\ref{eq:MeanEnergyChangeCommonForm}), (\ref{eq:MeanEnergyChangeCommonForm2}), and (\ref{eq:clockSubChanges}) indeed match the clock energy shift as \textit{originally} presented (Eq. (\ref{eq:clockMeanEnergyChange})).

\section{Rewriting the Clock Energy Shift}
Here we show that the clock mean energy shift with post-selection, $\Delta\langle{E_M}\rangle_f$, originally given as Eq. (\ref{eq:clockMeanEnergyChange}), indeed may be written in the form of Eq. (\ref{eq:MeanEnergyChangeCommonForm}) and (\ref{eq:MeanEnergyChangeCommonForm2}) (thus falling into the same general pattern as the Jaynes-Cummings result), with values substituted from Eq. (\ref{eq:clockSubChanges}). Note that Eq. (\ref{eq:MeanEnergyChangeCommonForm}) and (\ref{eq:MeanEnergyChangeCommonForm2}) are written in terms of energy quanta (dimensionless units). Given that $\hat{H}_0=\frac{\omega_0}{2}\hat{\sigma}_z$, the dimensionless form of Eq. (\ref{eq:clockMeanEnergyChange}) is:

\begin{equation}\label{appendixEq:clockMeanEnergyChangeRewrite}
\begin{split}
\frac{1}{\omega_0}\Delta\langle{E_M}\rangle_f
&\approx\frac{1}{2}\Re\left(\frac{\bra{f}\left(\ket{\uparrow_z}\bra{\uparrow_z}-\ket{\downarrow_z}\bra{\downarrow_z}\right)\ket{i}}{\bra{f}\ket{i}}\right)\\*
&-\frac{1}{2}\bra{f}\left(\ket{\uparrow_z}\bra{\uparrow_z}-\ket{\downarrow_z}\bra{\downarrow_z}\right)\ket{f}\\*
&=\Re\left(\frac{\bra{f}\ket{\uparrow_z}\bra{\uparrow_z}\ket{i}}{\bra{f}\ket{i}}\right)-1\\*
&\textcolor{green}{
\underbrace{\textcolor{black}{-\left(\bra{f}\ket{\uparrow_z}\bra{\uparrow_z}\ket{f}-1\right)}}_{\textcolor{black}{\mathrm{2nd \, drive \, term \, of \, Eq. (\ref{eq:MeanEnergyChangeCommonForm})}}
}}\\*
&=\Re\left(\frac{\bra{f}\ket{\uparrow_z}\bra{\uparrow_z}\ket{i}}{\bra{f}\ket{i}}\right)\textcolor{green}{
\underbrace{\textcolor{black}{-\bra{f}\ket{\uparrow_z}\bra{\uparrow_z}\ket{f}}}_{\textcolor{black}{\mathrm{2nd \, drive \, term \, of \, Eq. (\ref{eq:MeanEnergyChangeCommonForm2})}}}}
\end{split}
\end{equation}
Here we have used the completeness relation $\hat{1}=\ket{\uparrow_z}\bra{\uparrow_z}+\ket{\downarrow_z}\bra{\downarrow_z}$, which implies that $\frac{1}{2}\left(\ket{\uparrow_z}\bra{\uparrow_z}-\ket{\downarrow_z}\bra{\downarrow_z}\right)=\ket{\uparrow_z}\bra{\uparrow_z}-\frac{1}{2}$. We have explicitly included the second drive terms (green underbrace) found in Eq. (\ref{eq:MeanEnergyChangeCommonForm}) and (\ref{eq:MeanEnergyChangeCommonForm2}) to make for an easier comparison. To check Eq. (\ref{eq:MeanEnergyChangeCommonForm}) and (\ref{eq:clockSubChanges}), we expand out $\Re\left(\frac{\bra{f}\ket{\uparrow_z}\bra{\uparrow_z}\ket{i}}{\bra{f}\ket{i}}\right)-1$ and check that it matches the first term of Eq. (\ref{eq:MeanEnergyChangeCommonForm}). This is just algebra, and we include the steps here for completeness:

\begin{equation}
\begin{split}
&\Re\left(\frac{\bra{f}\ket{\uparrow_z}\bra{\uparrow_z}\ket{i}}{\bra{f}\ket{i}}\right)-1
=\frac{1}{2}\left(\frac{\bra{f}\ket{\uparrow_z}\bra{\uparrow_z}\ket{i}}{\bra{f}\ket{i}}+\frac{\bra{i}\ket{\uparrow_z}\bra{\uparrow_z}\ket{f}}{\bra{i}\ket{f}}\right)-1\\*
&=\frac{1}{2}\left(\frac{\bra{i}\ket{f}\bra{f}\ket{\uparrow_z}\bra{\uparrow_z}\ket{i}}{|\bra{f}\ket{i}|^2}+\frac{\bra{i}\ket{\uparrow_z}\bra{\uparrow_z}\ket{f}\bra{f}\ket{i}}{|\bra{f}\ket{i}|^2}\right)-\frac{\bra{i}\ket{f}\bra{f}\ket{i}}{|\bra{f}\ket{i}|^2}\\*
&=\frac{1}{|\bra{f}\ket{i}|^2}\Bigg(\frac{1}{2}\bra{i}\left(\ket{\uparrow_z}\bra{\uparrow_z}+\ket{\downarrow_z}\bra{\downarrow_z}\right)\ket{f}\bra{f}\ket{\uparrow_z}\bra{\uparrow_z}\ket{i}+\frac{1}{2}\bra{i}\ket{\uparrow_z}\bra{\uparrow_z}\ket{f}\bra{f}\left(\ket{\uparrow_z}\bra{\uparrow_z}+\ket{\downarrow_z}\bra{\downarrow_z}\right)\ket{i}\\*
&\qquad\qquad\quad-\bra{i}\left(\ket{\uparrow_z}\bra{\uparrow_z}+\ket{\downarrow_z}\bra{\downarrow_z}\right)\ket{f}\bra{f}\left(\ket{\uparrow_z}\bra{\uparrow_z}+\ket{\downarrow_z}\bra{\downarrow_z}\right)\ket{i}\Bigg)\\*
&=\frac{1}{|\bra{f}\ket{i}|^2}\Bigg((-1)\bra{i}\ket{\downarrow_z}\bra{\downarrow_z}\ket{f}\bra{f}\ket{\downarrow_z}\bra{\downarrow_z}\ket{i}+(-1/2)\left(\bra{i}\ket{\uparrow_z}\bra{\uparrow_z}\ket{f}\bra{f}\ket{\downarrow_z}\bra{\downarrow_z}\ket{i}+\overline{\bra{i}\ket{\uparrow_z}\bra{\uparrow_z}\ket{f}\bra{f}\ket{\downarrow_z}\bra{\downarrow_z}\ket{i}}\right)\Bigg)\\*
&=\frac{1}{|\bra{f}\ket{i}|^2}\Bigg((-1)|\bra{f}\ket{\downarrow_z}\bra{\downarrow_z}\ket{i}|^2+(-1/2)\cdot2\Re\left(\bra{f}\ket{\uparrow_z}\bra{\uparrow_z}\ket{i}\overline{\bra{f}\ket{\downarrow_z}\bra{\downarrow_z}\ket{i}}\right)\Bigg).
\end{split}
\end{equation}

This shows that the clock energy shift (Eq. (\ref{eq:clockMeanEnergyChange})) indeed follows the form of Eq. (\ref{eq:MeanEnergyChangeCommonForm}) with values substituted from Eq. (\ref{eq:clockSubChanges}). Analogous reasoning shows that the clock energy shift (Eq. (\ref{eq:clockMeanEnergyChange})) also follows the form of Eq. (\ref{eq:MeanEnergyChangeCommonForm2}).

This also clarifies that the measurement model using the \textit{degenerate} qubit-oscillator interaction gives the same results as the clock model. The photon number shift from the degenerate qubit-oscillator interaction model was already shown (in Appendix E) to follow Eq. (\ref{eq:MeanEnergyChangeCommonForm}) and (\ref{eq:MeanEnergyChangeCommonForm2}), with values substituted from Eq. (\ref{eq:clockSubChanges}).

\end{widetext}

\bibliography{ref.bib}

\end{document}